\documentclass{pasj00}

\begin{document}
\SetRunningHead{Bosch-Ramon \& Khangulyan}{Secondary Radio emission in $\gamma$-Ray Binaries}

\title{Monte-Carlo Simulations of Radio Emitting Secondaries in $\gamma$-Ray Binaries}

\author{Valenti \textsc{Bosch-Ramon} %
   \thanks{}}
\affil{Dublin Institute for Advanced Studies, Fitzwilliam Place 31, Dublin 2, Ireland}
\email{valenti@cp.dias.ie}

\author{Dmitry \textsc{Khangulyan}}
\affil{Institute of Space and Astronautical Science/JAXA, 3-1-1 Yoshinodai, Chuo-ku, Sagamihara, Kanagawa 252-5210, Japan}\email{khangul@astro.isas.jaxa.jp}

%


\maketitle

\begin{abstract}
Several binary systems that contain a massive star have been detected in both the radio band and at very high
energies.  In the dense stellar photon field of these sources,  gamma-ray absorption and pair creation are expected
to occur, and the radiation from these pairs  may contribute significantly to the observed radio emission. We aim at
going deeper in the study of the properties, and in particular the morphology, of the  pair radio emission in
gamma-ray binaries. We apply a Monte-Carlo code that computes the creation location, the spatial trajectory and the
energy evolution of the pairs produced in the binary system and its surroundings. The radio emission produced by
these pairs, with its spectral, variability and spatial characteristics, is calculated as it would be seen from a
certain direction. A generic case is studied first, and then the specific case of LS~5039 is also considered. We
find that, confirming previous results, the secondary radio emission should appear as an extended radio structure of
a few milliarcseconds size. This radiation would be relatively hard, with fluxes up to $\sim 10$~mJy. Modulation is
expected depending on the gamma-ray production luminosity, system eccentricity, and wind ionization fraction, and to 
a lesser extent on the magnetic field structure. In
gamma-ray binaries in general, the pairs created due to photon-photon interactions can contribute significantly to
the core, and generate an extended structure. In the case of LS~5039, the secondary radio emission is likely to be
a significant fraction of the detected core flux, with a marginal extension. 
\end{abstract}

\section{Introduction}

High-mass binary systems are a well established class of very high-energy (VHE) gamma-ray emitters. Five VHE
sources\footnote{There is also the new GeV gamma-ray binary, 1FGL~J1018.6$-$5856, which may have also been detected at VHE (e.g. \cite{cor11,deo10}).} 
have been associated so far to binaries that contain a massive star: PSR~B1259$-$63 
\citep{aha05a}; LS~5039 \citep{aha05a}; LS~I~+61~303 \citep{albert06}; \mbox{Cygnus~X-1} \citep{albert07}; and
HESS~J0632$+$057 \citep{aha07,hinton09,skilton09,falcone10,falcone11,mol11}. Among these five, three show clear modulation of the
VHE emission associated to the orbital motion: PSR~B1259$-$63 \citep{aha05a}, LS~5039 \citep{aha06}, and LS~I~+61~303
\citep{albert09}; thus, the emitter cannot be too far from the star in these sources. The short duration of the VHE flare
observed in \mbox{Cygnus~X-1} \citep{albert07} likely implies that this flare is originated not too far from the star,
whereas in HESS~J0632$+$057 the situation is less clear (see \cite{skilton09,falcone10,falcone11,mol11}). Under the strong
ultraviolet photon field of a massive star, gamma-ray absorption can take place  (e.g.
\cite{ford84,protheroe87,moskalenko94,bed00,boettcher05,dubus06a,boettcher07,orellana07,khangulyan08, rey08,sier08,rom10}),
this process leading to the creation of secondary (electron-positron) pairs with energies $\gtrsim 10$~GeV and peaking around
30~GeV. 

Secondary pairs are created under the stellar photon field and, as
shown below, are likely trapped by the stellar wind magnetic field of
strength $B_{\rm w}$. For small enough values of $B_{\rm w}$ and high
enough values of the secondary particle energy $E$, most of the
pair energy is emitted, via inverse Compton (IC) scattering of stellar
photons, as gamma-rays with energies above the pair-creation
threshold. If the gamma-ray opacity coefficient $\tau_{\gamma\gamma}$
is $\gg 1$, the electromagnetic cascade may significantly affect the
distribution of secondary pairs in the system. However, for this to be
efficient the initial gamma-ray energy must be well above the pair
creation energy threshold, $\epsilon\gg\epsilon_{\rm th}\sim
1/\epsilon_*$ (i.e. deep in the Klein-Nishina -KN- regime), where
$\epsilon_{\rm th}$, $\epsilon$ and $\epsilon_*$ are the threshold, the gamma-ray and the stellar photon
energies in $m_{\rm e}\,c^2$ units. In this regime, say for
energies of the secondary pairs $\gtrsim 1$~TeV and typical stellar
temperatures $\sim (3-4)\times 10^4$~K, synchrotron cooling suppresses
the cascade under a magnetic field \citep{khangulyan08}
\begin{equation} 
B_{\rm w}\gtrsim 10\,\left(L_*/10^{39}~{\rm erg~s}^{-1}\right)\left(R/10^{12}{\rm
cm}\right)^{-1}\,{\rm G}\,,
\label{bkn} 
\end{equation} 
where $L_*$ and $R$ are the star luminosity and distance from the star center,
respectively. Equation~\ref{bkn} has been derived from Eq.~1 in \citet{khangulyan08}, which gives 
the IC 
cooling timescale in the KN regime with a 10~eV photon field as target. We reproduce 
the formula here for
the sake of clarity:
\begin{equation}
t_{\rm KN}\approx 1.7\times 10^2\,(u_*/100\,{\rm erg~cm}^{-3})^{-1}\,(E/1\,{\rm TeV})^{0.7}\,{\rm s}\,,
\end{equation}
where $u_*$ is the stellar radiation energy density, and $E$ is the electron energy.
For $B_{\rm w}>\sqrt{8\,\pi\,u_*}$, the
secondary bolometric luminosity output will be dominated by synchrotron radiation. This implies a magnetic field
strength: 
\begin{equation} 
B_{\rm w}\gtrsim 3\times 10^2\,(L_*/10^{39}{\rm erg~s}^{-1})^{1/2}(R/10^{12}{\rm
cm})^{-1}\,{\rm G}\,. 
\end{equation} 
All this shows that even for magnetic fields well below equipartition with radiation,
synchrotron losses can effectively prevent development of electromagnetic cascades. 

\citet{bosch08a} carried out a detailed semi-analytical study of the spectral properties of the secondary broadband
emission for realistic magnetic field values (see also \cite{bosch08b}). In that work, it was shown that the radio,
X-ray, and GeV fluxes of secondary emission could be comparable to those observed in gamma-ray binaries, and may
even dominate the non-thermal output below $\epsilon_{\rm th}$ in some cases. These authors also suggested that the
extended radio emission found in LS~I~+61~303 \citep{dhawan06} may come from secondary pairs. \citet{bosch09a} also
proposed this origin for the emission of the marginally resolved radio core of LS~5039 \citep{ribo08}.

Although the radio fluxes can be roughly estimated from the semi-analytical study of \citet{bosch08a}, their
approach does not allow a proper description of the radio spectrum, lightcurve, and morphology, because synchrotron
radio emitting particles have long timescales and are strongly affected by the medium inhomogeneity at spatial
scales of $\sim R$. In that work, the complex 3-dimensional (3D) structure of the secondary trajectories was
simplified attaching the secondary pairs to the stellar wind, with the latter being taken as spherically symmetric,
with constant velocity, and wind particle and magnetic field energy density $\propto 1/R^2$. The magnetic field was
considered fully irregular. Finally, the adiabatic cooling of the secondary pairs was not taken into account, which
is reasonable at binary spatial scales, but not for the computation of that radio emission
produced outside the binary system, affected by wind expansion. Finally, the relevance of free-free
absorption was not investigated. 

\begin{figure*}
\center{}
\includegraphics[width=0.4\textwidth]{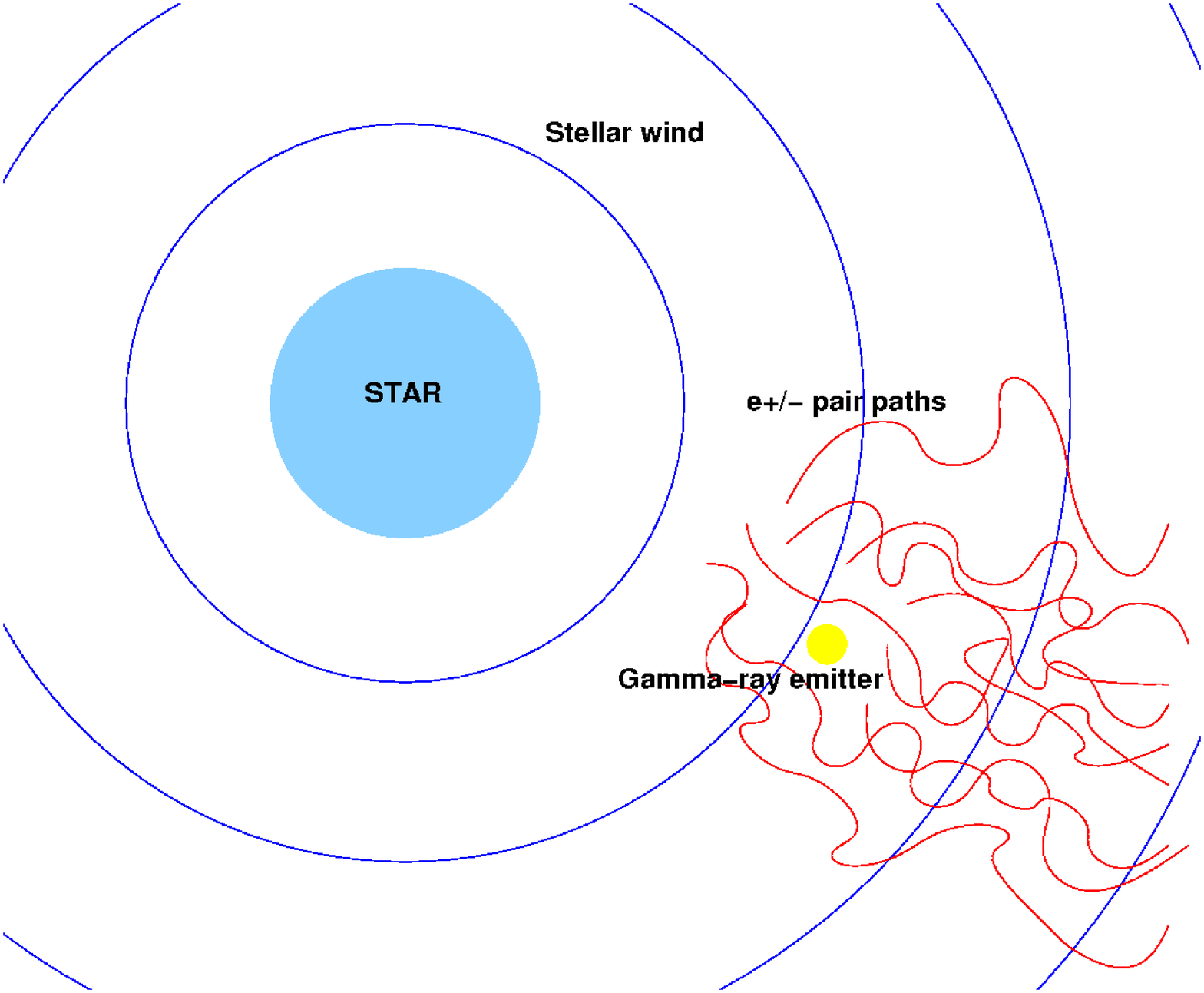}\qquad
\includegraphics[width=0.4\textwidth]{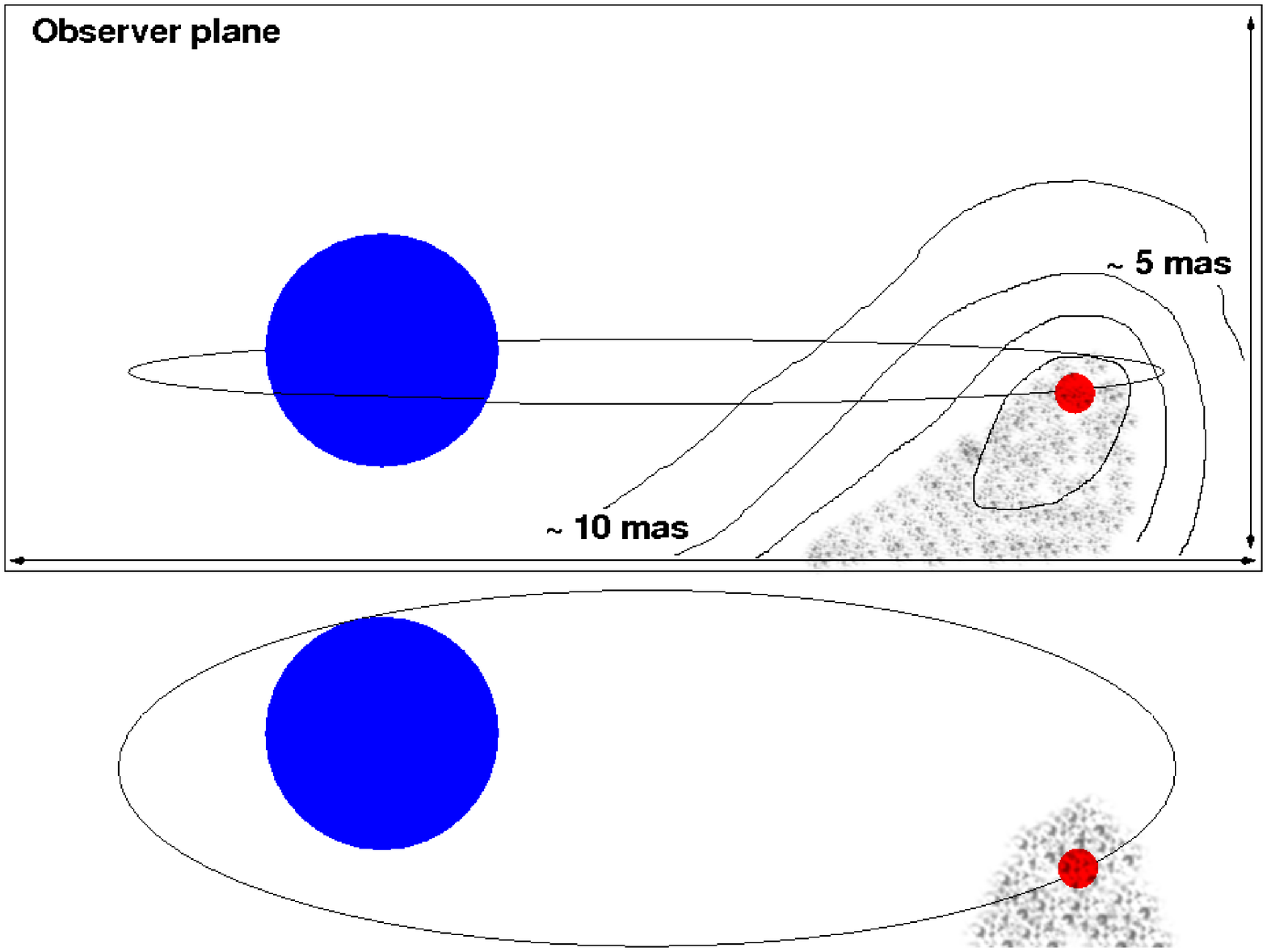}\\[10pt]
\caption{Left: Sketch of the considered scenario at the scales of the binary system, { looking at the system 
perpendicularly to the orbital plane}. Secondary pairs are created
in the vicinity  of the gamma-ray emitter, get trapped in the stellar wind through slow diffusion, and are
advected by it away from the star. Right: Sketch of the radio emission produced by secondary pairs as seen by the
observer. The radiation emitted by the advected secondary pairs, of spiral shape due to wind advection and orbital
motion, will appear  smoothed by the radio telescope beam.}
\label{fig1}
\end{figure*}

In this work, we present detailed calculations of the evolution of the radio emitting secondary pairs in gamma-ray
binaries. The energy and the spatial evolution of the particles in the stellar wind have been computed using a
Monte-Carlo code that accounts for ionization/coulombian collision losses, adiabatic, relativistic bremsstrahlung,
synchrotron, and IC cooling,  as well as diffusion and advection in the wind to follow the trajectories of secondary
pairs in the source. The code can also deal with anisotropic diffusion when an ordered magnetic field is dominant,
although we do not consider this configuration at this stage. A sketch of the the scenario at the scales of the
binary system is shown in Fig.~\ref{fig1}. Once the particle energy and spatial distributions have been obtained at
a particular orbital phase { (accounting for the previous orbital history of the system)}, the radio emission in a
specific direction is calculated, which allows the generation of directional radio maps.  The radio maps have been
smoothed using a circular Gaussian in order to roughly reproduce what an observer would see using a milliarcsecond
(mas) resolution radio interferometer. This is also illustrated in Fig.~\ref{fig1}.  Spectra and lightcurves, and
the impact of free-free absorption, have been discussed. We focus here on a generic case, since our aim is to give a
general demonstration of the relevance of secondary radio emission, but we provide also with an instance of the importance of this
phenomenon in a real source, applying our calculations to LS~5039. Further and more detailed studies for specific
objects of the role of secondary pairs for the radio emission will be presented elsewhere.

\section{Monte-Carlo simulations of secondary pairs in gamma-ray binaries}\label{phys}

For an appropriate study of the secondary radio emission, detailed calculations of the energy and spatial evolution  of
the secondary pairs are required, accounting for the star radiation field and the properties of the stellar wind. Given
the difficulties of treating the magnetic field structure and adiabatic cooling in the wind in a semi-analytical
approach, we have performed Monte-Carlo simulations in which secondaries pairs are injected, move (diffuse and are
advected), and cool down in a 3D space.The 3D nature of the problem comes from the combination of the 
advection in the wind plus the orbital motion (a significant regular 
$B_{\rm w}$-field structure would also require a 3D treatment).
The problem treated by the Monte-Carlo method 
is similar to that described by the diffusion advection equation 
(e.g. \cite{jones90,bla87}).
The radiation timescales ($t=|E/\dot{E}|$, where $\dot E$ is the electron energy loss rate) adapted to the environments of gamma-ray binaries 
are:
\begin{equation}
  t_{\rm sync}\approx 40\,(B/100~{\rm G})^{-2}\,(E/1~{\rm GeV})^{-1}\,{\rm s}\,,
\end{equation}
\begin{equation}
t_{\rm IC~Th}\approx 50\,(u_*/300~{\rm erg~cm}^{-3})^{-1}\,(E/1~{\rm GeV})^{-1}\,{\rm s}\,,
\end{equation}
\begin{equation}
t_{\rm br}\approx 10^5\,(n_{\rm w}/10^{10}\,{\rm cm}^{-3})^{-1}\,{\rm s}\,,
\end{equation}
for synchrotron, IC (Thomson) and relativistic Bremsstrahlung, respectively (see \cite{bosch09b} for the IC cooling rate 
accounting also for the KN regime used in the calculations), 
with $n_{\rm w}$ being the wind density (see below), $u_*$ the energy density of the target photon field, and $B$ the ambient magnetic field.
For
the adiabatic and ionization cooling timescales, we have adopted:
$$
t_{\rm ad}=(3/2)\,(R/v_{\rm w\infty})\,(1-R_*/2R)^{-1}, {\rm which~for}\,R\gg R_*
$$
\begin{equation}
\approx 2\times 10^4\,(R/3\times 10^{12}\,{\rm cm})\,
(v_{\rm w\infty}/2\times 10^8\,{\rm cm~s}^{-1})^{-1}\,{\rm s}\,,
\end{equation}
and
\begin{eqnarray}
t_{\rm ion}&\approx& 2\times 10^{18}\,E/n_{\rm w}
\nonumber\\ {} & \approx &
3.4\times 10^6\,(E/1~{\rm GeV})\,(n_{\rm w}/10^9\,{\rm cm}^{-3})^{-1}\,{\rm s}\,,
\end{eqnarray}
respectively, where $v_{\rm w\infty}$ is the wind velocity at infinity, and $R_*$ the stellar radius.

\subsection{Injection of secondary pairs in the binary system}\label{phys}

We have calculated the injected spatial and energy distribution of secondary pairs for a given primary gamma-ray
spectrum and a stellar black-body photon field. This is done tracking the path of a photon produced in the emitter
location and moving with a certain direction in the binary system.  
The location in which a given gamma ray is absorbed is obtained using the { (anisotropic) gamma-ray absorption differential 
probability} due to photon-photon interactions (point-like treatment of the star; see below):
\begin{equation}
\frac{{\rm d}w}{{\rm d}l}=\left(1-\cos\theta\right)
\int{\rm d}\epsilon_*
\sigma_{\gamma\gamma}\left(\epsilon\epsilon_*\left(1-\cos\theta\right)\right)
n_*(\epsilon_*,R)\,,
\end{equation}
where $l$, $\theta$, $n_*$ and $\sigma_{\gamma\gamma}$ are the length of the path covered by the gamma ray,
the gamma-ray/stellar photon interaction angle, the stellar
photon specific density, and the pair-creation cross section \citep{gould67}, respectively.
This quantity is integrated along the photon path $\int_{0}^l {\rm d}w$ to obtain the probabity
of interaction up to a certain location.
In this work, the absorption probability is computed along the
photon trajectory with small steps, and the Monte-Carlo algorithm is applied to define the interaction point.

The injection rate of secondary pairs depends strongly on the spectrum and angular distribution of the primary
gamma rays.  To mimic the effect of the anisotropic IC scattering in the angular distribution of primary gamma
rays, a gamma-ray direction probability at injection $\propto (1-\cos\theta_*)$ has been introduced, where
$\theta_*$ is the angle between the gamma-ray path and the direction from the star to the gamma-ray emitter, i.e.
gamma rays have a higher probability to  be emitted towards than away from the star.  This angular dependence
would roughly correspond to an average one between the Thomson and KN IC regimes (see Fig.~6
in \cite{khangulyan08}).

\begin{figure*}
\centering
\includegraphics[width=1\textwidth]{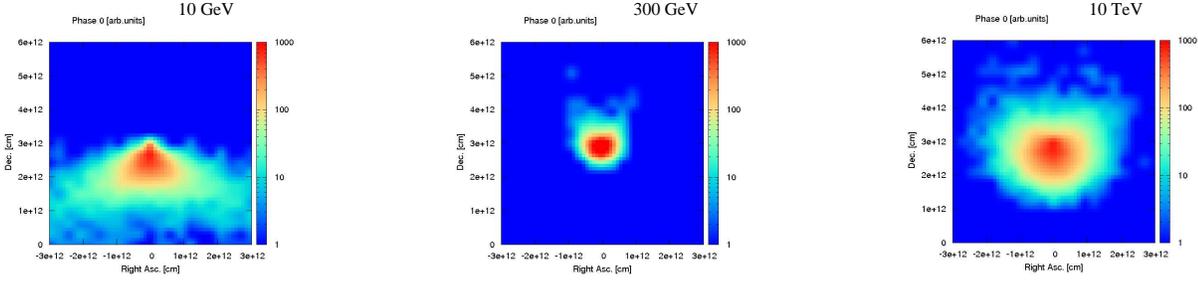}
\caption{Projection in the orbital plane of the spatial distribution 
of the secondary pairs injected at three different energies: $E=10$~GeV (left panel), 300~GeV (central panel) 
and 10~TeV (right panel). 
This has been computed using Monte-Carlo simulations.
The gamma-ray emitter is located at $(x,y)=(0,3\times 10^{12})$~cm, and the 
star at $(x,y)=(0,0)$~cm. Most of the secondary pairs are produced towards the star, with a
plateau in their concentration 
up to the distance at which $\tau_{\gamma\gamma}\sim 1$. This distance changes with energy and has a minimum around
few 100~GeV for most of the interaction angles, corresponding to a peak in the pair creation cross section. 
Intensity units are arbitrary.}\label{fig0}
\end{figure*}

In Fig.~\ref{fig0}, we show the distribution of injected secondary pairs in a gamma-ray binary for three different
energies. We have focused here on a generic case, in which we have adopted a circular binary system ($e=0$) of one
week period, formed by a massive, primary star of mass $M_*$ and compact object of mass $M_{\rm X}$, such that
$M_*+M_{\rm X}=22$~$M_{\odot}$. This yields an orbital separation distance of $R_{\rm orb}=3\times 10^{12}$~cm. The
inclination of the system has been fixed to $i=45^\circ$, and the phase 0.0/0.5 has been taken at the
superior/inferior conjunction of the compact object (supc/infc). The main star parameters have been fixed to
$L_*=6\times 10^{38}$~erg~s$^{-1}$, $R_*=10^{12}$~cm and $T_*=3\times 10^4$~K, where $T_*$ is the star temperature.
The gamma-ray emitter has been assumed to be point-like, at the compact object location, with $R_{\rm e}=R_{\rm
orb}$, although the actual location and size of the VHE emitter in gamma-ray binaries has not been defined yet,
being possible to have an extended production region not only in the jet but also in the colliding wind scenario
(see, e.g., \cite{bogovalov08}). No eccentricity and a fixed $i$-value have been taken for the sake of simplicity,
since our aim is to illustrate the main characteristics of the secondary radio emitter. 

\subsubsection{Treatment of the star}

\begin{figure*}
\centering
\includegraphics[width=0.65\textwidth]{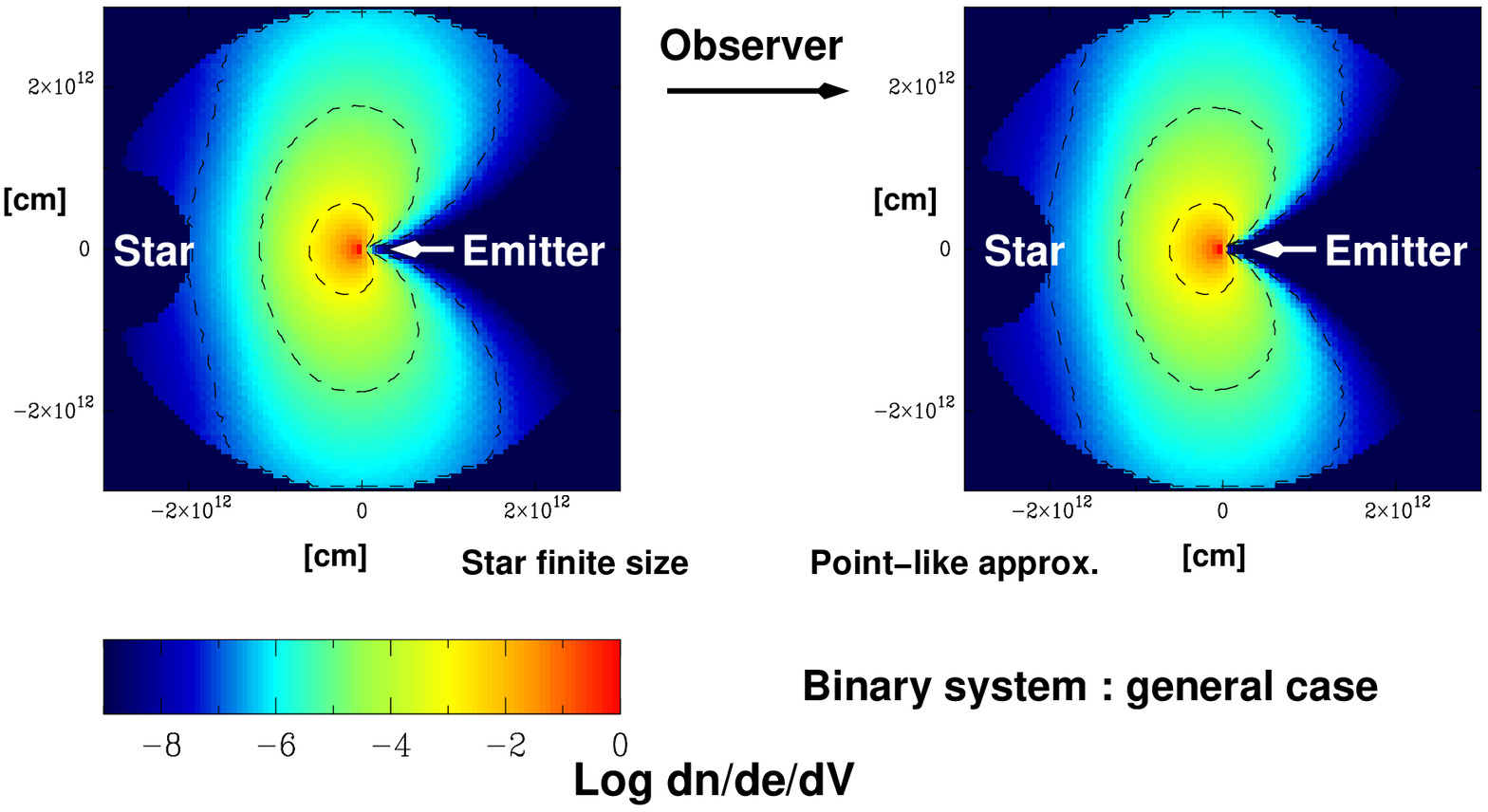}\\[10pt]
\includegraphics[width=0.65\textwidth]{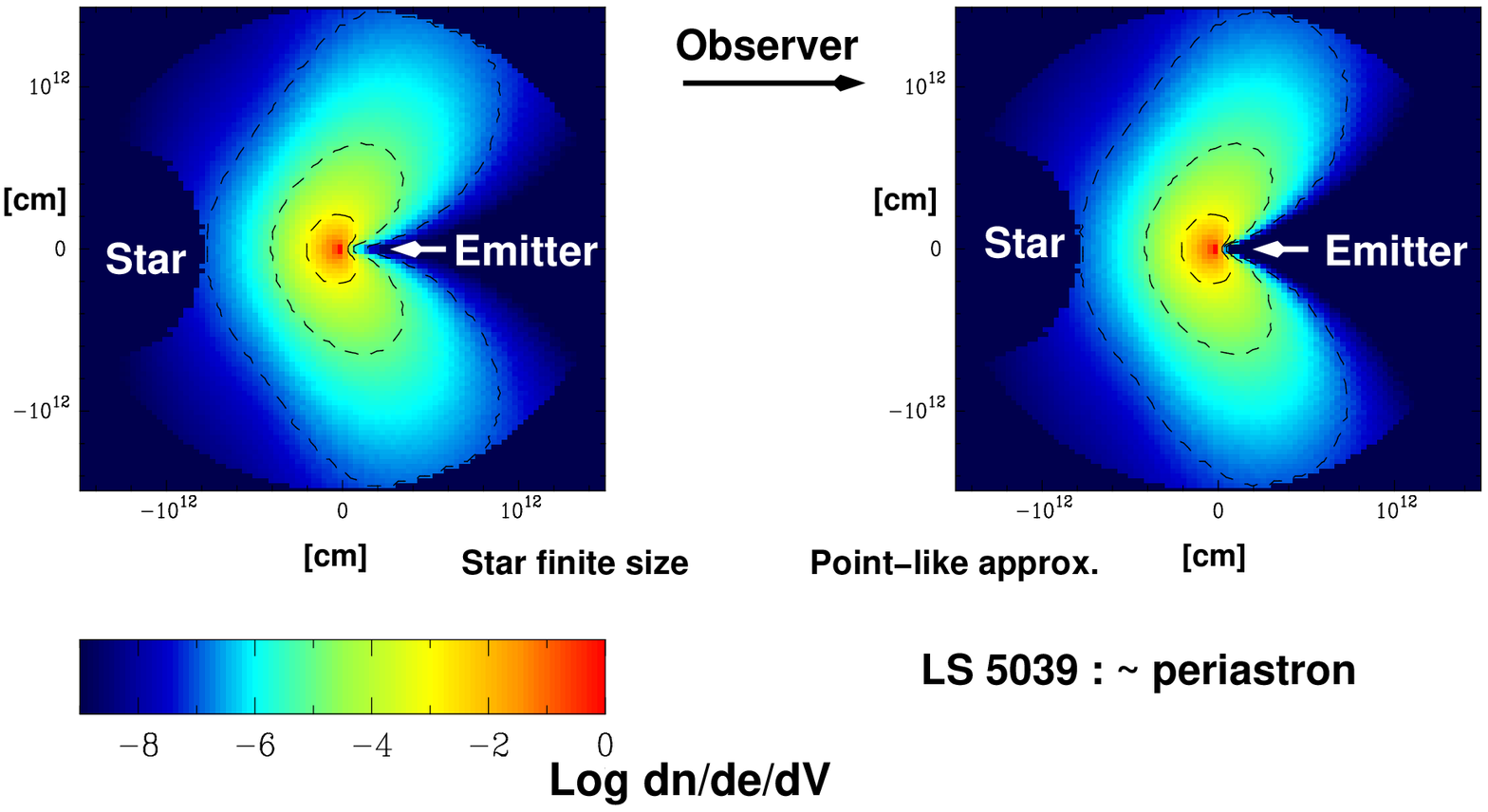}
\caption{Maps of the injection rate per volume unit for injected pairs of 100~GeV, dominant in radio. This has been computed 
semi-analytically.
Two cases are presented: 
a generic case (top) with the parameter values of Table~1, and a case with the properties of LS~5039 
around periastron (bottom; see Table~2). The calculation accounting for the 
star finite size is shown at the left, and the point-like approximation at the right. Units are the logarithm of the rate density 
normalized to the peak value. The 
star is located to the left in both maps.}
\label{mapin}
\end{figure*}

\begin{figure}
\centering
\includegraphics[width=0.38\textwidth]{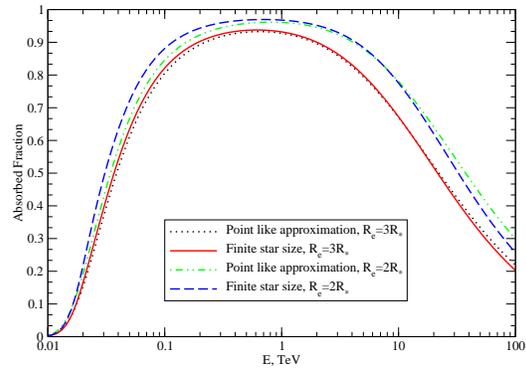}
\caption{Fraction of the primary gamma-rays absorbed in the stellar
photon field. The gamma-ray emitter was
assumed to have a $(1-\cos\theta)$-angular dependence (see Sect.~2.1
for details). 
The calculations were performed for the star
properties given in Table~1.
The dotted line corresponds to the point-like star approximation and a star-gamma-ray emitter distance of $R_{\rm e}=3R_*$; 
the solid line corresponds to the stellar finite size case and a star-gamma-ray emitter distance of $R_{\rm e}=3R_*$; 
the dash-dot-dot line corresponds to the point-like star approximation and a star-gamma-ray emitter distance of $R_{\rm e}=2R_*$; 
the dashed line corresponds to the stellar finite size case and a star-gamma-ray emitter distance of $R_{\rm e}=2R_*$.}
\label{sec_integ}
\end{figure}

In these calculations, we have adopted a point-like approximation for
the star (e.g. \cite{dubus06a}; see also Sect.~4.2.1 in
\cite{bosch09b}). This approximation fails close to the stellar
surface, at a distance comparable to the radius of the star
$R_*$, i.e. $R\ltsim 2\,R_*$. This region will be relevant for gamma
rays with pair-production mean-free paths comparable to the distance
$R_{\rm e}$ between the star and the gamma-ray emitter.  Such gamma
rays can be absorbed in a volume $\sim 4\pi\,R_{\rm e}^3/3$, with a
sub-region affected by the star finite-size of volume
$\sim \pi(2\,R_*)^2\times R_*$. Thus, even for the most sensitive interval
of gamma-ray energies, the point-like approximation of the stellar
photon field fails for a fraction of the total relevant volume $\sim
3\,(R_*/R_{\rm e})^3\sim 0.11$ and $0.38$ for $R_{\rm e}=3\,R_*$ and
$2\,R_*$, respectively. Therefore, given a rather broad distribution
of primary gamma rays and for $R_{\rm e}\gtrsim 2\,R_*$, the
point-like approximation for the star would have an accuracy at least
of the order of a 10\%. This estimate is suported by semi-analytical
calculations of the pair injection rate density, whose results are
  presented in Fig.~\ref{mapin}. The calculation is done computing
  first the gamma-ray flux that reaches a certain point in the binary
  system. Then, the rate density of pairs of 100~GeV (dominant in radio) injected
  in each point is computed, accounting for the energy differential
  cross section of pair creation (see \cite{bosch08a} and references
  therein). This calculation have been performed for a generic case and for LS~5039 (see Table~2).  
 In Fig.~\ref{sec_integ}, we show the fraction of primary gamma-rays
absorbed in the system, as a function of the gamma-ray energy, for two
different distances between the gamma-ray emitter and the
star: $R_{\rm e}=3\,R_*$ and $R_{\rm e}=2\,R_*$, with the properties of the
star chosen as in the generic case.  
As seen from these figures, accounting 
the stellar finite size or adopting a point-like star
approximation yield almost the same results.
This permits to
safely adopt the point-like approximation for the Monte-Carlo
calculations presented below, shortening significantly the calculation
time.

\subsubsection{Primary gamma-ray spectrum and flux}

\begin{figure}
\centering
\includegraphics[width=0.38\textwidth]{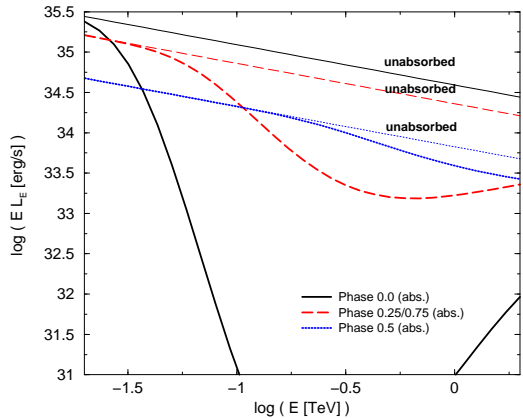}
\caption{Primary gamma-ray emission between 20~GeV and 2~TeV in the direction to the observer, for four different orbital phases: 
0.0 (solid line), 0.25/0.75 (long-dashed line), and 0.5 (dotted line). The thin lines show the unabsorbed spectra, 
and the thick ones show the same but accounting for photon-photon absorption.}\label{figg}
\end{figure}

To perform the calculations corresponding to a generic gamma-ray binary, the photon index of the produced gamma
rays has been taken $\Gamma=2.5$, and the unabsorbed differential photon rate at 1~TeV is $N_{\rm TeV}=1.4\times
10^{34}$~TeV$^{-1}$~s$^{-1}$, which corresponds to a specific photon flux $n_{\rm TeV}=3\times
10^{-11}$~TeV$^{-1}$~s$^{-1}$~cm$^{-2}$ at a distance of $d=2$~kpc. These values are assumed to be orbital phase independent. 
In Fig.~\ref{figg}, we show the primary gamma-ray emission between 20~GeV to 2~TeV in the observer direction, for
the orbital phases 0.0 (supc), 0.25, 0.5 (infc), and 0.75. 
The thin lines correspond to the intrinsic spectra, whereas the thick ones take into account photon-photon absorption.
The computed absorbed specific photon
fluxes are, at 1~TeV: $n_{\rm TeV}\sim 10^{-14}$ (phase 0.0), $2\times 10^{-12}$ (phase 0.25/0.75) and $5\times
10^{-12}$~TeV$^{-1}$~s$^{-1}$~cm$^{-2}$ (phase 0.5).

\subsection{Energy and spatial evolution of secondary pairs}\label{phys}

\subsubsection{Transport of secondary pairs}

A secondary pair is injected where the gamma ray was absorbed, and the subsequent locations are
computed accounting for the local direction and strength of the wind magnetic field, which is assumed to have an
ordered and a disordered component. We consider the stellar rotation axis to be perpendicular to the orbital plane.

The ordered wind magnetic field can be described
as ${\bf B_{\rm w}}(r,\phi)=B_{\rm r}{\bf \hat e_{\rm r}}+B_\phi{\bf \hat e_\phi}$, where 
\begin{equation}
B_{\rm r}\approx B_*(R_*/R)^2, 
\end{equation}
\begin{equation}
B_\phi\approx B_*(v_{\rm
w\phi}/v_{\rm w\infty})(R_*/R)\,,
\end{equation} 
and $B_*$ and $v_{\rm w\phi}$ are the magnetic field at the star surface and 
the initial azimuthal wind velocity, respectively. For the calculations, $B_*$ is fixed
to a value of 200~G (see \cite{bosch08b} and references therein), high enough to suppress the
electromagnetic cascade. We neglect here the region
within the Alfven radius from the star, typically confined to a region $R_{\rm A}<2\,R_*$, since almost all 
secondary pairs are created farther from the star. 
The wind velocity can be expressed as 
${\bf v_{\rm w}}(r,\phi)=v_{\rm wr}{\bf \hat e_{\rm r}}+v_{\rm w\phi}{\bf \hat e_\phi}$, where:
\begin{equation}
v_{\rm wr}(R)\approx v_{\rm w\infty}\,(1-R_*/R)\,,
\end{equation}
and the azimuthal one:
\begin{equation}
v_{\rm w\phi}(R)\approx 0.1\,v_{\rm w\infty}\,(R_*/R)\,.
\end{equation}
The wind density depends on $R$ as 
\begin{equation}
n_{\rm w}(R)=\dot{M}_{\rm w}/4\,\pi\,R^2\,v_{\rm wr}(R)\,,
\end{equation}
where $\dot{M}_{\rm w}$ is the star mass-loss rate.
In this work, we have
taken $\dot{M}_{\rm w}=10^{-6}\,M_{\odot}$~yr$^{-1}$ and $v_{\rm w\infty}=2.5\times 10^8$~cm~s$^{-1}$. 
The magnetic field and the wind model  adopted here are based on that
presented in \citet{usov92}.

The irregular component of the magnetic field, $\delta B$, allows particles to effectively diffuse in the stellar wind. The
particle mean free paths parallel and perpendicular to the magnetic lines are $\lambda_{\parallel}=\eta\,r_{\rm g}$ and
$\lambda_\perp=r_{\rm g}/\eta$, where $r_{\rm g}=3\times10^8 (E/100{\rm \, GeV})(B_{\rm w}/1{\rm \,G})^{-1}\rm \, cm$ is the electron gyro radius, and $\eta$ is $\ge 1$ and relates the
irregular and the wind magnetic field component through $B_{\rm w}=\sqrt{\eta}\,\delta B$. 
(thus the ordered component strength is $\sqrt{\eta-1}\delta B$).
The spatial power spectrum of
$\delta B$ is assumed to be flat, i.e. $\eta$ does not depend on energy, and the diffusion coefficients  are proportional to
the Bohm one: $\kappa_{\parallel,\perp}=\lambda_{\parallel,\perp}\,c/3$.

The diffusion approximation is applicable if 
secondary pairs deflect significantly before cooling down, i.e. $t_{\parallel}=\lambda_\parallel/c\ll t_{\rm cool}$:
\begin{equation}
B_{\rm w}\gg 10^{-3}\,(\eta/10^3)\,(E/{\rm 1~GeV})\,(t_{\rm cool}/10^2\,{\rm cm~s}^{-1})^{-1}\,{\rm G}\,.
\end{equation}
In addition, the particle mean free path parallel to the magnetic lines (always longer than the perpendicular one)
should be much smaller than the system typical scale $R$:
\begin{eqnarray}
\lambda_\parallel/R&\approx& 10^{-3}\,(E/{\rm 1~GeV})\,(B_{\rm w}/1\,{\rm G})^{-1}\,
\nonumber\\ {} & \times &
(\eta/10^3)^{-1}\,(R/3\times 10^{12})^{-1}\,.
\label{lam}
\end{eqnarray}
The simulation time steps have to fulfill as well these conditions and in addition be $\gg t_{\parallel}$ for
the diffusion approximation to be suitable. All this has been accounted for in the calculations. 

The diffusion in the stellar wind has been accounted in each step $j$ of the secondary evolution as a location
shift, $\Delta \bf{X}$$_{\rm diff}^j=\Delta \bf{X}$$_{\parallel}^j\bf{\hat e_{\parallel}}$$+
\Delta \bf{X}$$_{\perp}^j\bf{\hat e_{\perp}}$, which corresponds to the diffusion displacement vector of the particle after a
time interval $dt^j$, accounting for the diffussion coefficients $\kappa_{\perp,\parallel}$. Since particles are
confined to the stellar wind, they will be also advected by the wind.  This advection motion has been accounted
for introducing an additional shift in the particle location after each time step $j$ by $\Delta\bf{X}$$_{\rm
adv}^j=\bf{v_{\rm w}}$$dt^j$. The total spatial shift of the secondary location per time step is therefore 
$\Delta\bf{X}$$_{\rm diff}^j+\Delta\bf{X}$$_{\rm adv}^j$.

For the sake of simplicity, at this stage we have assumed a fully disordered magnetic field with $\eta=1$. We note
that, as long as $\lambda_{\parallel}\ll R$, i.e.  for $\eta\ll 10^6$ (see Eq.~\ref{lam}), radio emitting
secondary pairs will basically move anchored to the stellar wind. In Table~1, the parameter values are given for
the generic case studied in this work.

 \begin{table}[]
  \begin{center}
  \caption[]{Binary system properties}
  \label{tab0}
  \begin{tabular}{lll}
  \hline\noalign{\smallskip}
  \hline\noalign{\smallskip}
Parameter [units] & Symbol  &  Value   \\
  \hline\noalign{\smallskip}
Stellar radius [cm] & $R_*$ & $10^{12}$ \\
Orbit separation distance [cm] & $R_{\rm orb}$ & $3\times 10^{12}$ \\
Emitter-to-star distance [cm] & $R_{\rm e}$ & $3\times 10^{12}$ \\
Stellar temperature [K] & $T_*$ & $3\times 10^4$ \\
total system mass [$M_{\odot}$] & $M_*+M_{\rm X}$ & 22 \\
Star surface magnetic field [G] & $B_*$ & 200 \\
Mass-loss rate [$M_{\odot}$~yr$^{-1}$] & $\dot{M}$ & $10^{-6}$ \\
Wind speed at infinity [cm~s$^{-1}$] & $v_{\infty}$ & $2.5\times 10^8$ \\
Distance [kpc] & $d$ & 2 \\
Superior conjunction & supc & 0.0 \\
Inferior conjunction & infc & 0.5 \\
Inclination angle [$^\circ$] & $i$ & 45 \\
Irregular magnetic field fraction & $\eta$ & 1 \\
1~TeV specific flux [TeV$^{-1}$~s$^{-1}$] & $n_{\rm TeV}$ & $3\times 10^{-11}$ \\
Gamma-ray photon index & $\Gamma$ & 2.5 \\
  \noalign{\smallskip}\hline
  \end{tabular}
  \end{center}
\end{table}

\subsubsection{Secondary particle energy evolution}

Once created at time $t_\ast$, the secondary pair trajectory is
  calculated by the Monte-Carlo code up to time  $t$, i.e. the
  time when the energy and spatial distributions of pairs and
  their radiation are computed. The energy evolution of secondary pairs is
  calculated accounting for the energy losses along the trajectory
  taking into account all the processes in play: ionization,
  adiabatic, relativistic bremsstrahlung, synchrotron, and IC
  cooling. Thus, at each time step $j$, a certain amount of energy
$\Delta E^j=\dot{E}\,dt^j$ is rested to the particle energy at $j-1$,
where $dt^j$ is such that $\Delta E^j\ll E^{j-1}$, and $\dot{E}$
includes the cooling mechanisms specified above. We note that the
  energy losses depend not only on the particle energy, but also
  on the particle position in the system.

In the region close to the star ($R\sim$ few $R_*$), IC cooling
dominates for basically all the particle
energies, although at this distance there are none or very few radio
emitting particles. Farther out, adiabatic cooling in the wind becomes
faster. The energy
distribution of radio secondaries almost does not depend on the
original gamma-ray spectrum, since radio emitting secondary pairs are
mostly produced around $\epsilon_{\rm th}$ and then cool down.  
Dominant IC cooling yields an electron energy distribution below 
the pair creation threshold $\propto E^{-2}$, whereas 
adiabatic cooling renders $\propto E^{-1}$. 

In the Monte-Carlo simulation we have considered an equal number
  of primary gamma-rays ($10^4$ and $2\times 10^4$ for the generic and
  LS~5039 cases, respectively) per logarithmic energy bin with the
  angular distribution described in Sect.~2.1. The
  injection process has been implemented during the relevant time
  interval, starting early enough before the time $t$, to account for
  all the relevant secondary pairs (say, at $t_*\in [t_{\rm *min},t)$).  
  The orbital motion of the primary
  gamma-ray source has been accounted in the calculations of the
  injection rate.  The Monte-Carlo code calculates the secondary pair
  energy and space distribution at the time $t$ keeping record of the time
  of creation $t_\ast$ and the initial pair energy.  To compute the
  final radiative output of the secondary pairs one has to account for
  the adopted gamma-ray injection spectrum ($\Gamma=2.5$) and the
  corresponding normalization, i.e. one needs to know how many
  particles correspond in reality to one particle of the simulation.

  An analytical code has been used to compute the total amount of
  secondary pairs generated per energy bin for the primary gamma-ray
  distribution $dN_{\rm i\gamma}/d\gamma$. This value, divided by the
  number of secondary particles per energy bin injected in the simulation,
  gives the {\it weight} of an electron/positron from the evolved
  distribution of particles. This {\it weight} depends on the
  secondary particle energy, and may vary with the orbital phase,
  i.e. with injection time $t_\ast$.

  Pairs created at different epochs $t_\ast$ can contribute to the
  radio emission at the same frequency at the final time $t$, since
  pairs may have different initial energies and spatial trajectories
  and therefore cooling times differ as well.

\begin{figure*}
\centering
\includegraphics[width=0.3\textwidth]{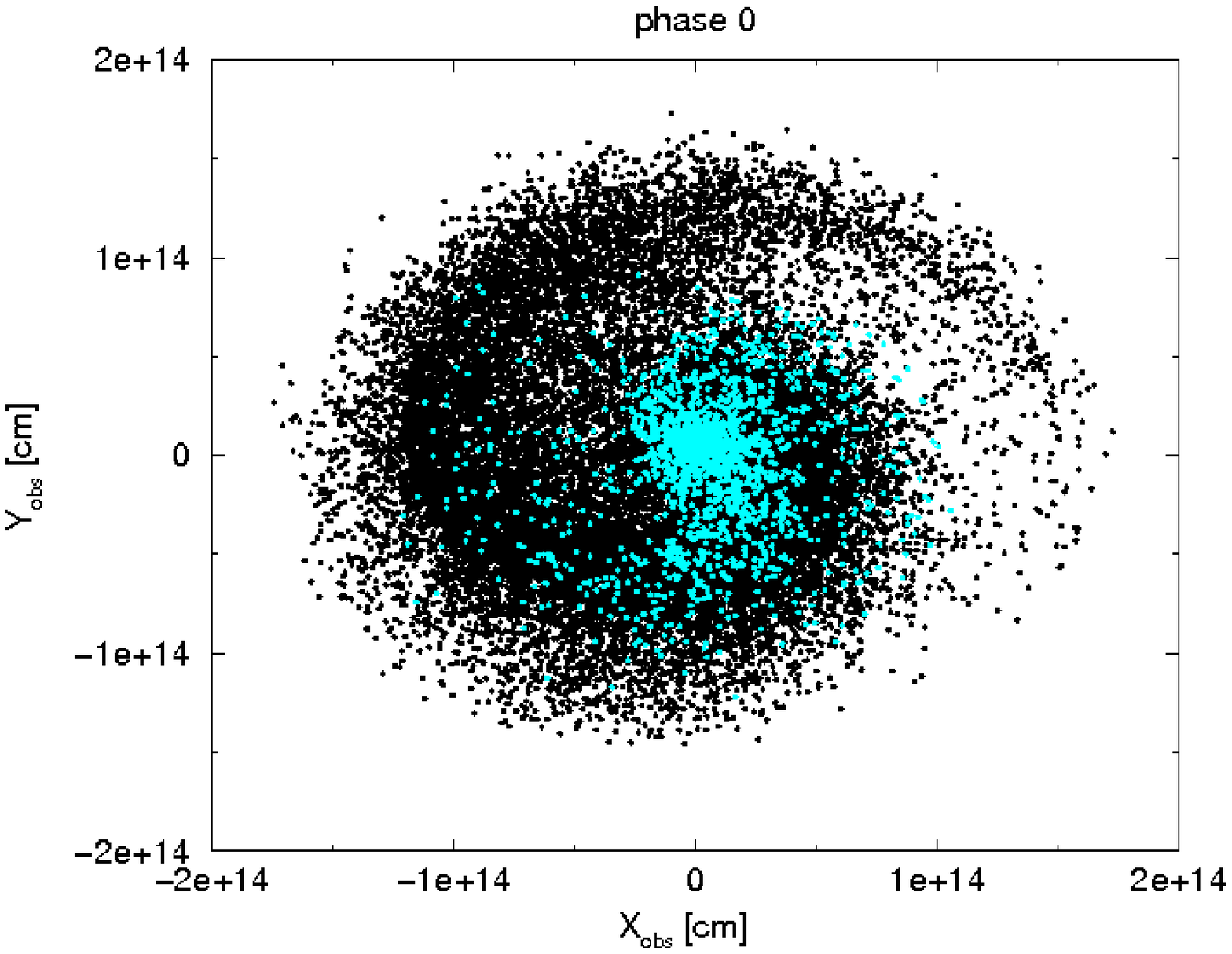}\qquad
\includegraphics[width=0.3\textwidth]{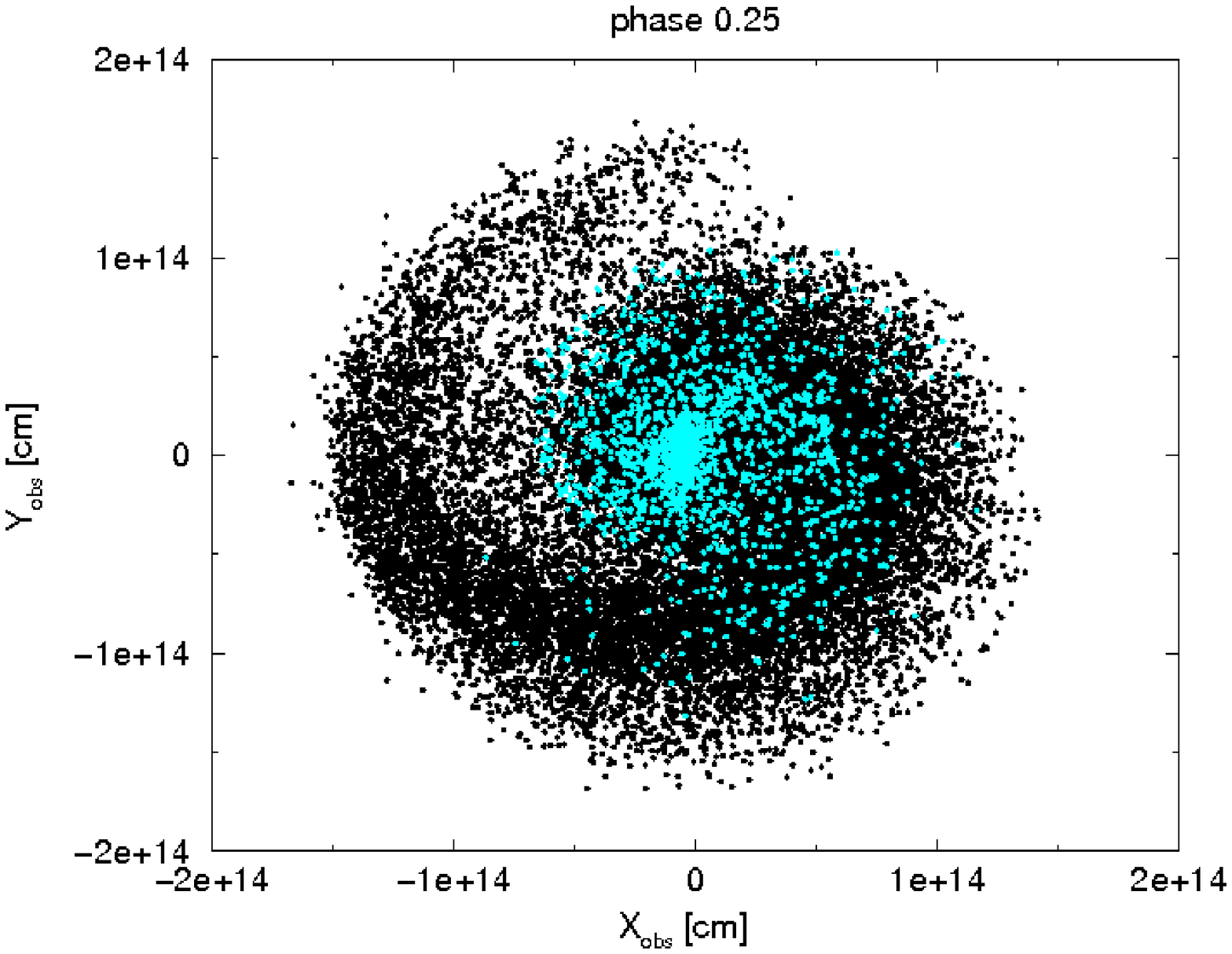}\\[10pt]
\includegraphics[width=0.3\textwidth]{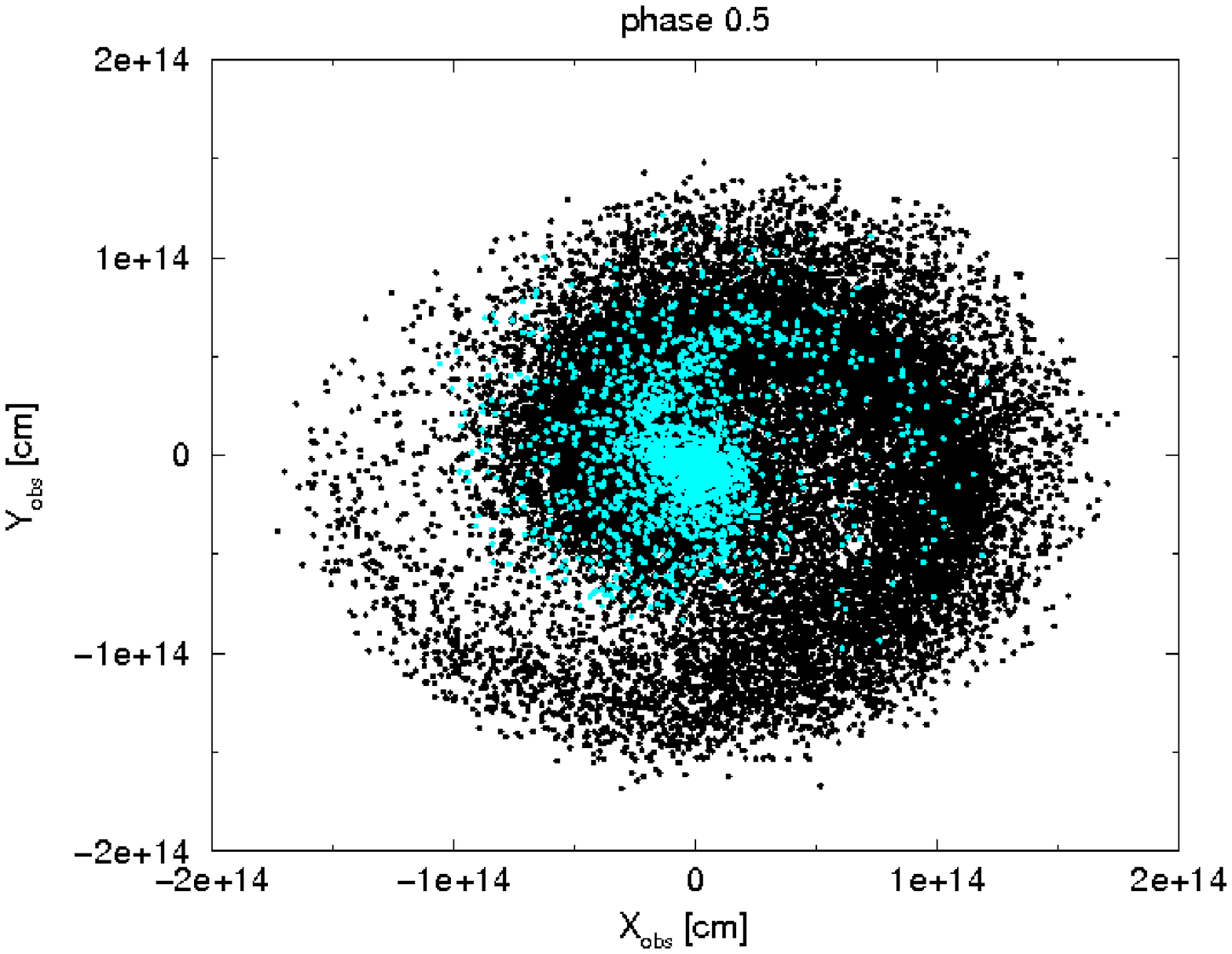}\qquad
\includegraphics[width=0.3\textwidth]{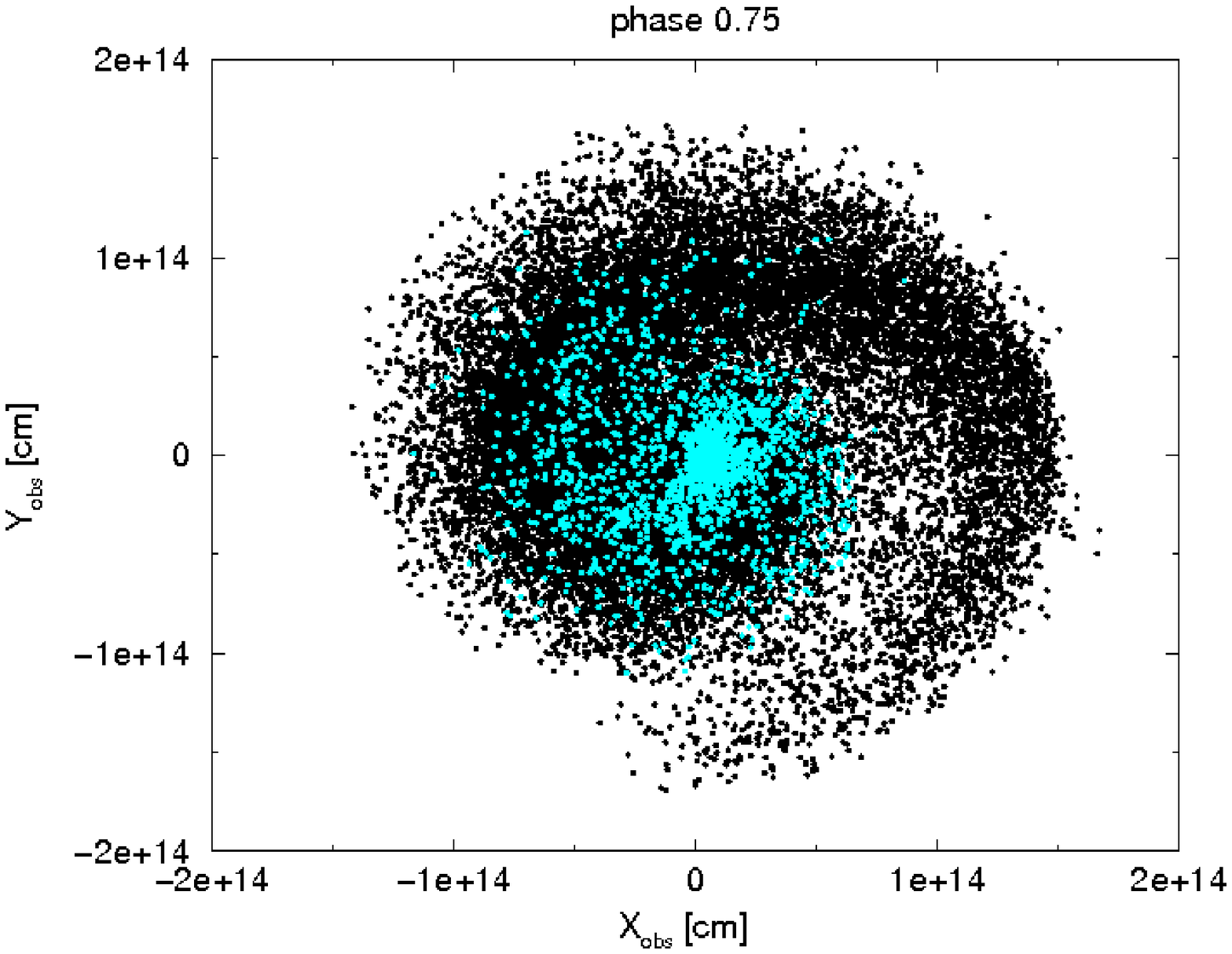}\\
\caption{Computed spatial distribution of the secondary pairs, projected in the observer plane 
($i=45^\circ$), for four different phases: 0.0 (top, left), 0.25 (top, right), 0.5 (bottom, left), 
and 0.75 (bottom, right). About $10^5$ particles have been injected (black spots),
among which about $10^4$ particles have GHz synchrotron emitting energies (light blue spots).}
\label{evol1}
\end{figure*}

\begin{figure*}
\centering
\includegraphics[width=0.35\textwidth]{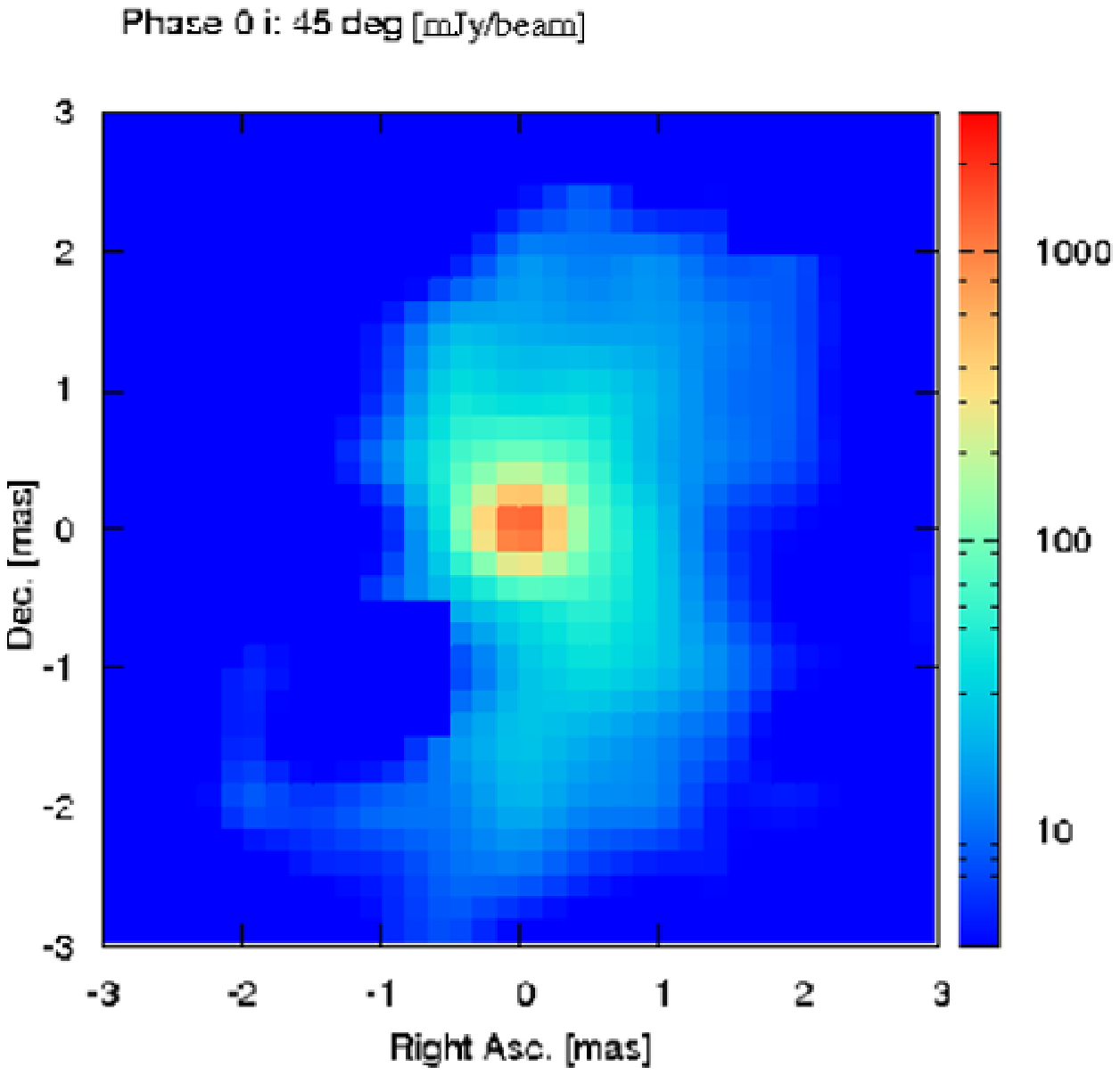}\qquad
\includegraphics[width=0.35\textwidth]{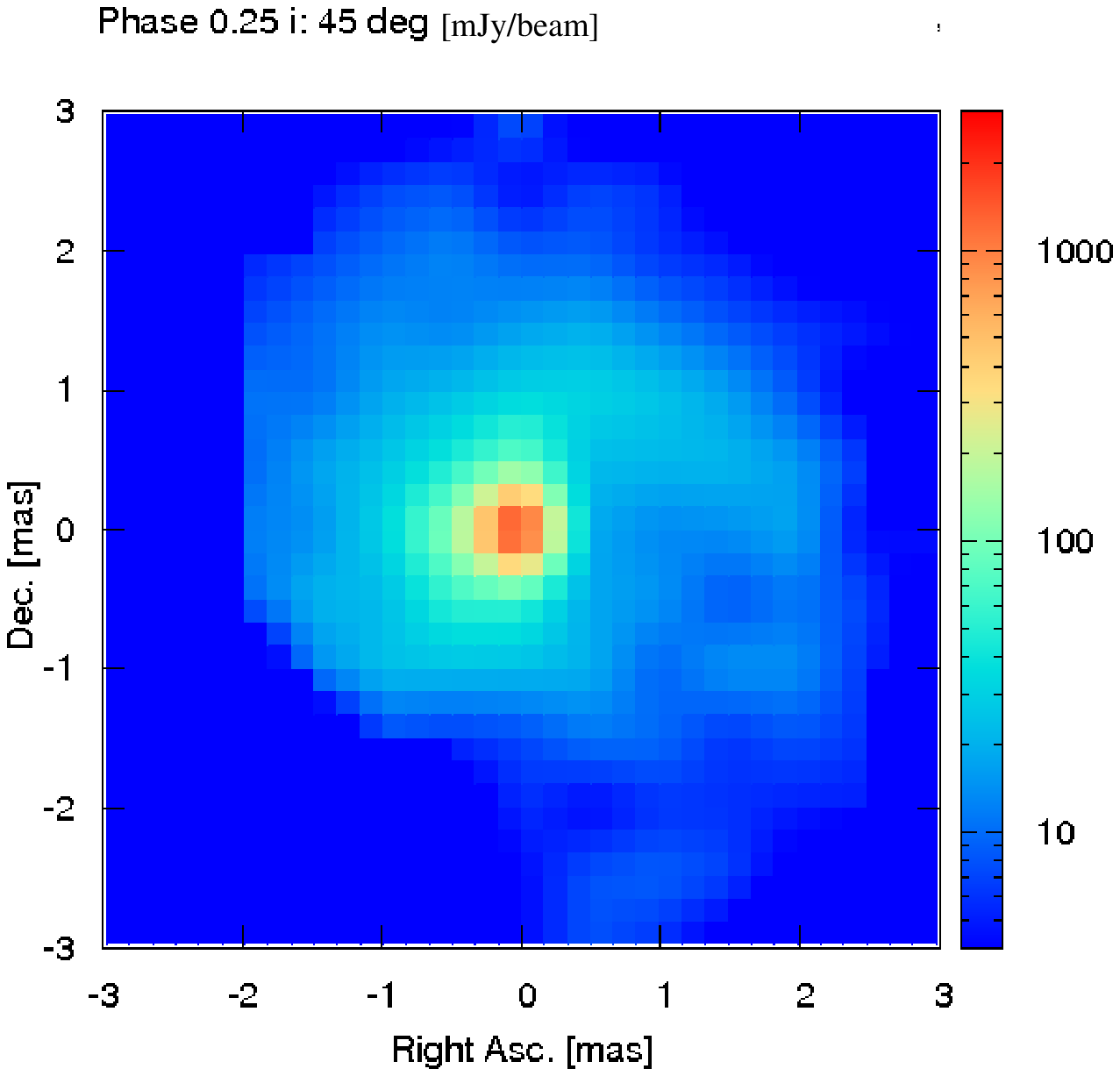}\\[10pt]
\includegraphics[width=0.35\textwidth]{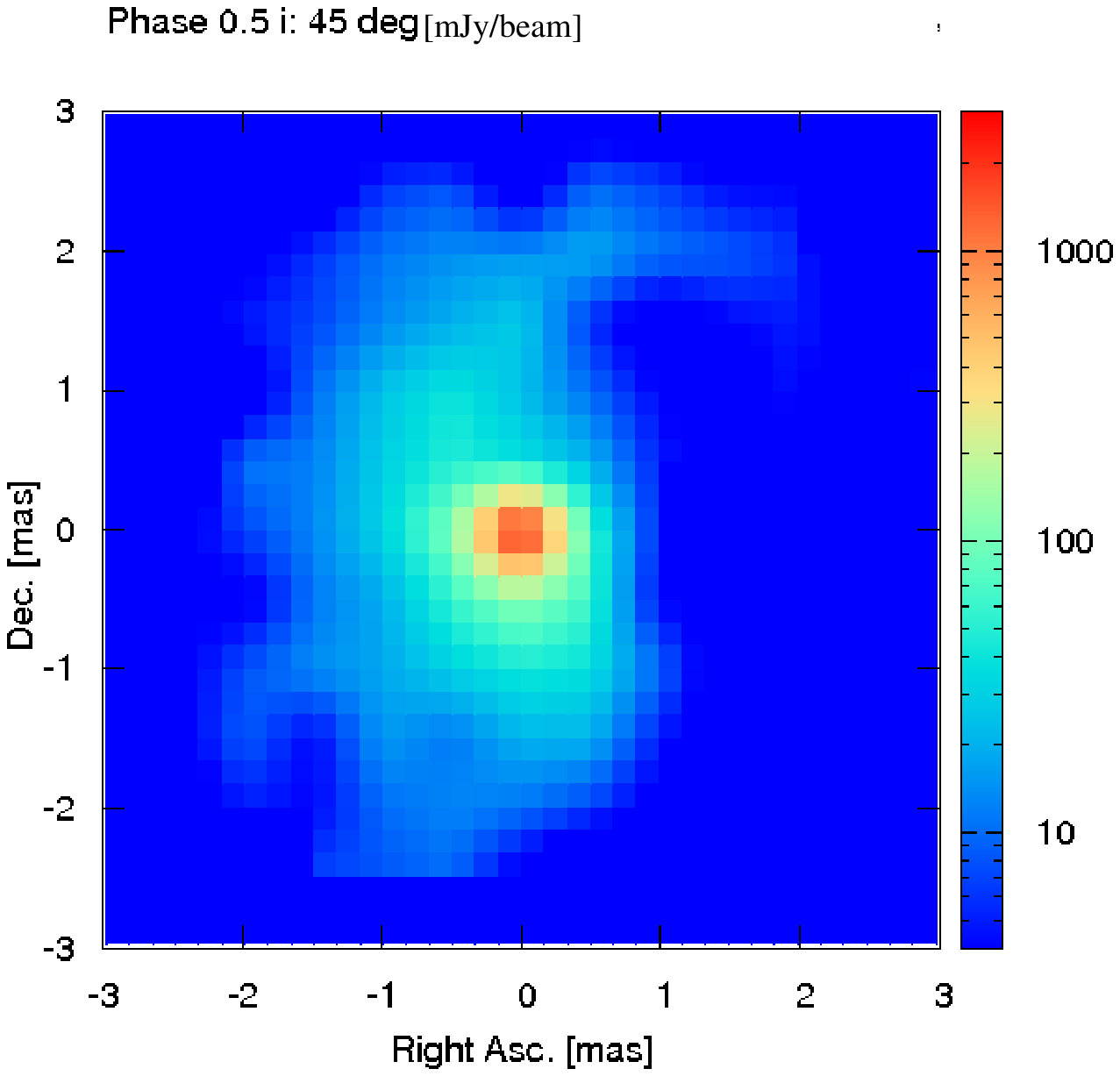}\qquad
\includegraphics[width=0.35\textwidth]{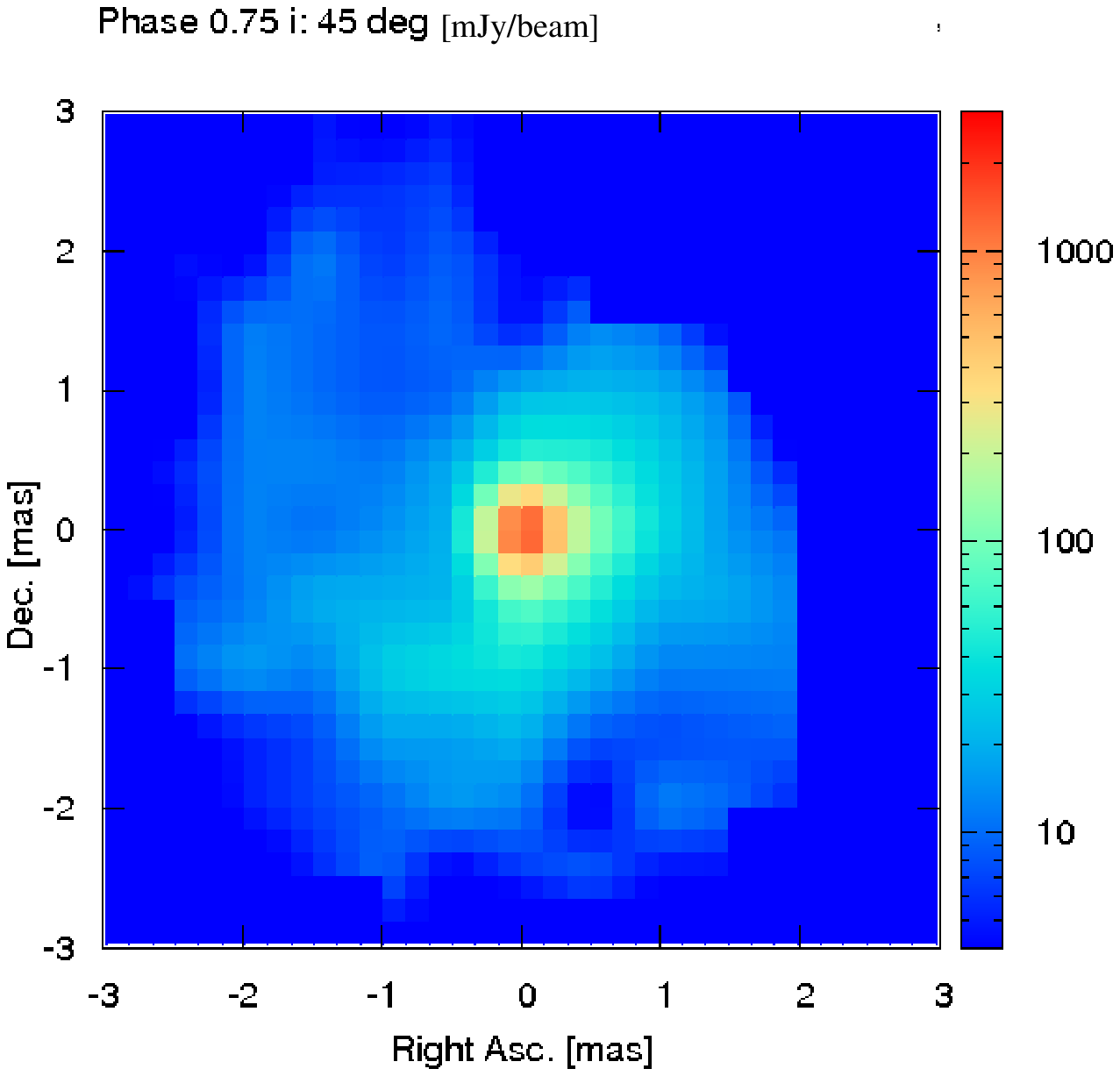}\\
\caption{Density map of the
radio secondary populations shown in Fig.~\ref{evol1}. Intensity units are arbitrary.}
\label{evol2}
\end{figure*}

In Figs.~\ref{evol1} and \ref{evol2}, we show the (evolved) secondary location distribution and density map,
respectively, projected in the observer plane for four different phases, 0.0, 0.25, 0.5, and 0.75.

\subsection{Secondary pair radiation}\label{phys}

From the energy and the spatial distribution of particles at a specific orbital phase, we have computed the
synchrotron emission convolving the particle energy distribution with the one-particle power function. Computing
the fluxes from 2D small regions or bins in the observer plane (image pixels), i.e. integrating in depth, permits to
extract radio morphologies. The radio emission has been calculated in the optically thin approximation of
synchrotron radiation, valid at frequencies above few GHz for the relevant spatial scales. For a 10~mJy source
with the self-absorption turnover at $\sim 1$~GHz, the radio emitter size would be constrained to $\approx 5\times
10^{13}\,(B/1~{\rm G})^{1/4}$~cm \citep{bosch09a} (where $B$ would be the radio emitter characteristic magnetic
field), or $\sim 1$~mas at few kpc.

\begin{figure}
\centering 
\includegraphics[width=0.3\textwidth]{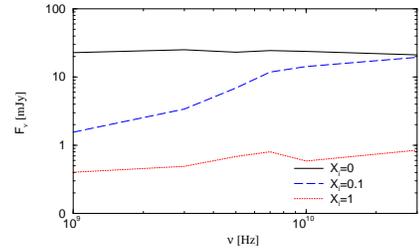} \caption{Computed spectra of the whole radio emitter
for three ionization fractions, $X_{\rm ion}=0$ (solid line), 0.1 (long-dashed line) and 1 (dotted line).}\label{spc}
\end{figure}

Typical radio fluxes of about 20~mJy are obtained at 5~GHz for the parameter values adopted here.  Note that, fixing 
$\Gamma$, the
secondary radio flux is $\propto N_{\rm TeV}/d^2$. The spectral shape of the specific
flux is almost flat, $F_\nu\propto$~constant, since adiabatic losses dominate the secondary pair evolution. Part of
the radio emission originates nevertheless in regions in which IC cooling dominates, which softens the spectrum.
Free-free absorption, of coefficient $\tau_{\rm ff\nu}$, could be relevant. Adopting a simple prescription for
$\tau_{\rm ff\nu}$ which does not depend on the orbital phase in a circular orbit (see Eq.~1 in  \cite{bosch09a},
adapted from \cite{ryb79}), we find that for ionization fractions $X_{\rm ion}\gtrsim 0.1$, the radio fluxes should be
already significantly affected. Another effect of the free-free absorption is a strong spectral hardening, although
for $X_{\rm ion}\rightarrow 1$ a combination of the energy and the spatial distribution of secondaries yields again
a flat spectrum. All this is shown in Fig.~\ref{spc}. Regarding the magnetic field geometry, fixing $\eta=1$ implies
that the synchrotron power does not depend on the magnetic field orientation and therefore the orbital phase. For
$\eta>1$, changes in the $B_{\rm w}$-line of sight angle could introduce moderate orbital flux variability. For a
dominantly toroidal magnetic field, synchrotron flux 
variations would be of the order of $\sin^2(\pi/2)/\sin^2(i)$ (2 for $i=45^\circ$), with
two dips and two enhancements along the orbit when the $B_{\rm w}$-line of sight angle reached
$i$ and $\pi/2$, respectively.

\begin{figure*}
\centering
\includegraphics[width=0.35\textwidth]{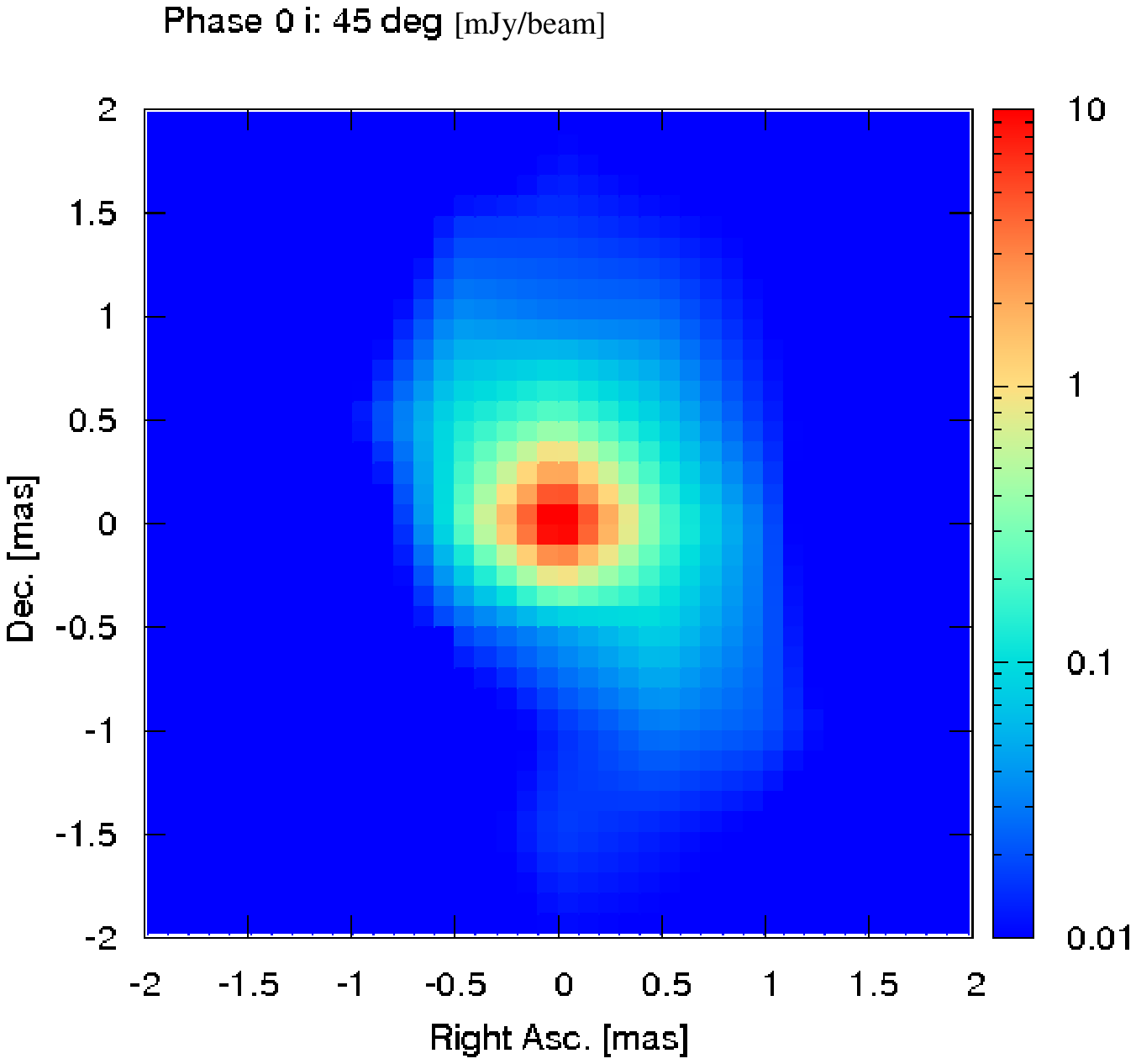}\qquad
\includegraphics[width=0.35\textwidth]{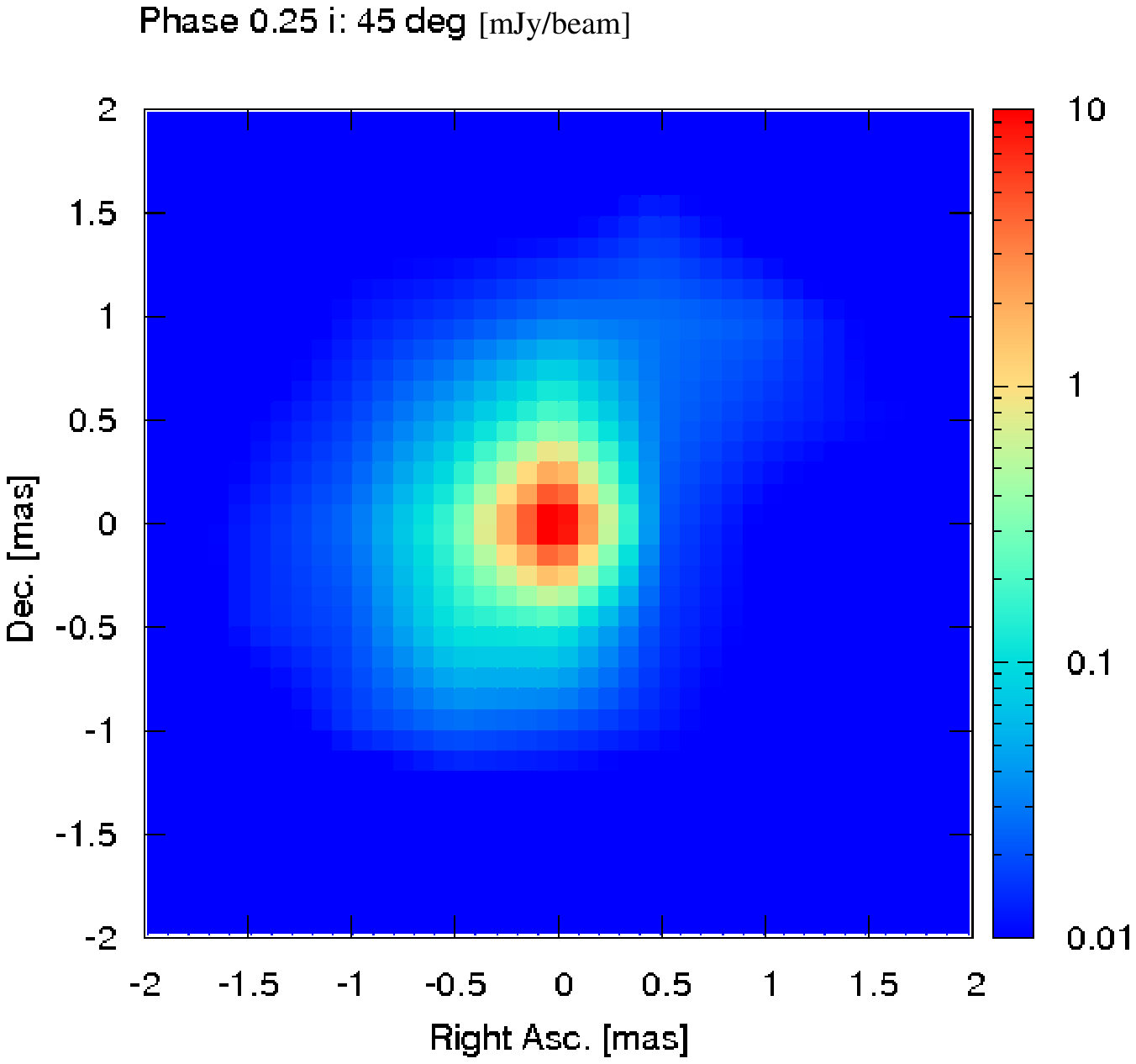}\\[10pt]
\includegraphics[width=0.35\textwidth]{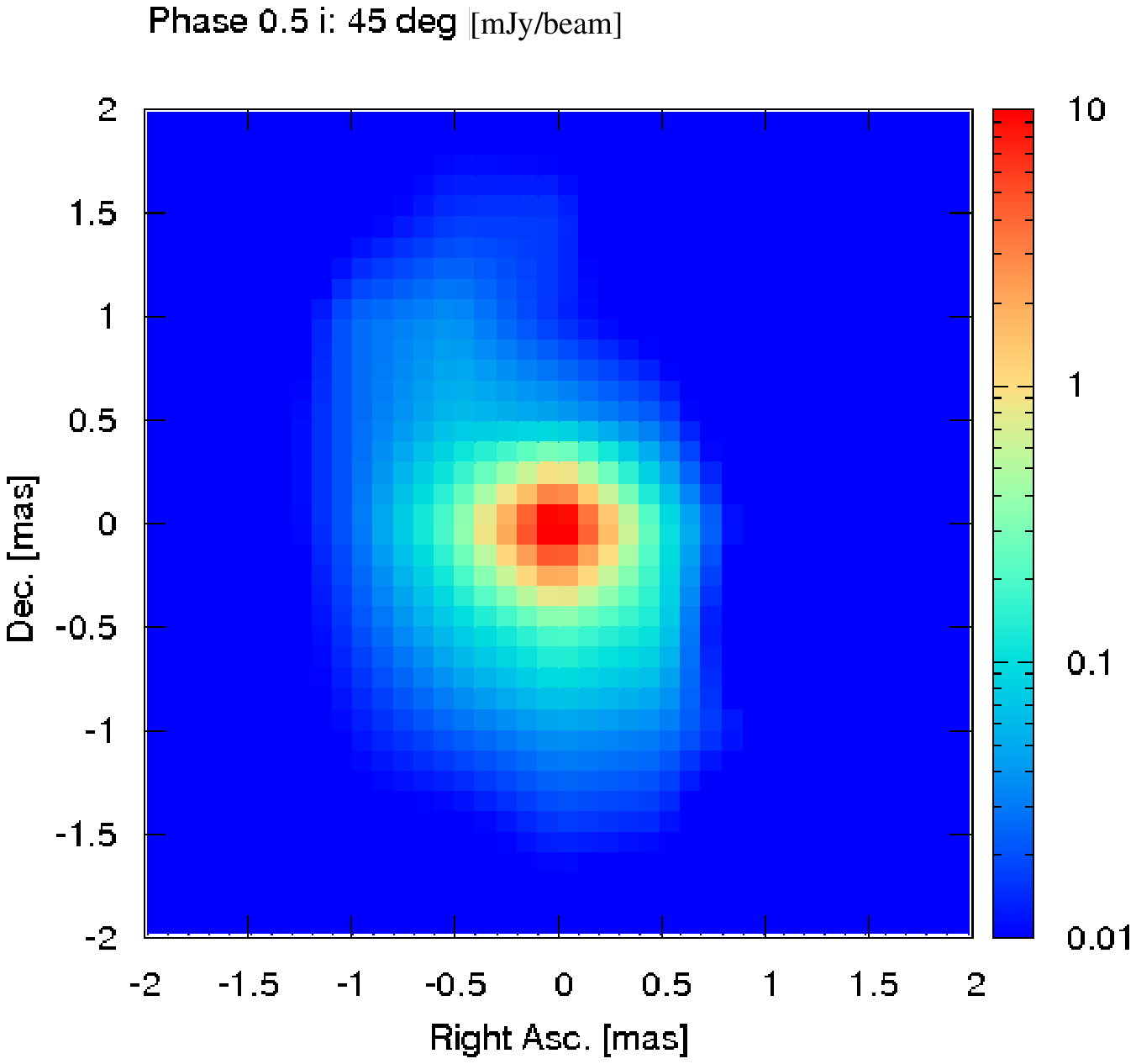}\qquad
\includegraphics[width=0.35\textwidth]{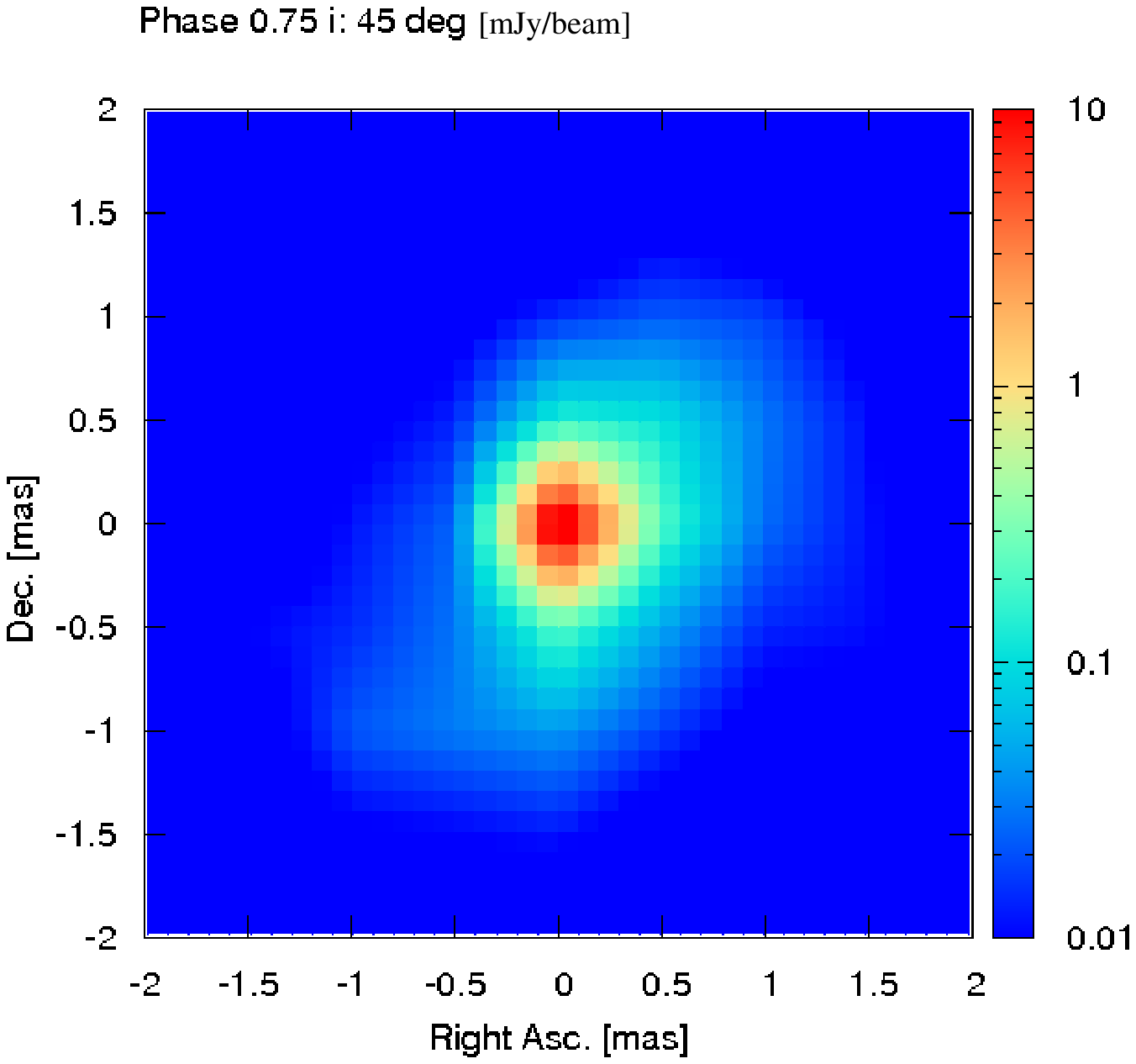}\\
\caption{Computed image of the 5~GHz radio emission, in the direction to the observer, in units of mJy per beam.}
\label{raw}
\end{figure*}

\begin{figure*}
\centering
\includegraphics[width=0.35\textwidth]{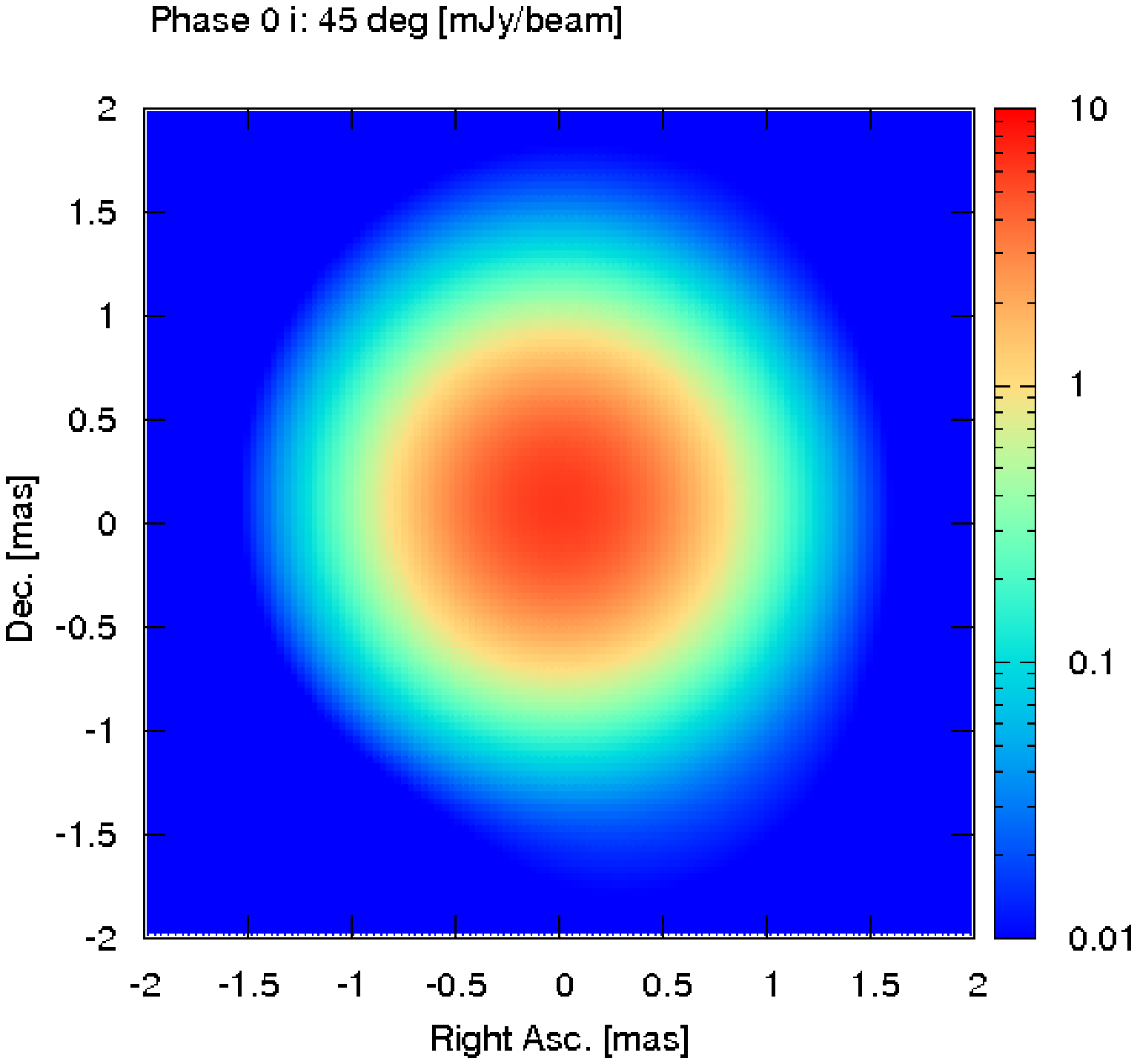}\qquad
\includegraphics[width=0.35\textwidth]{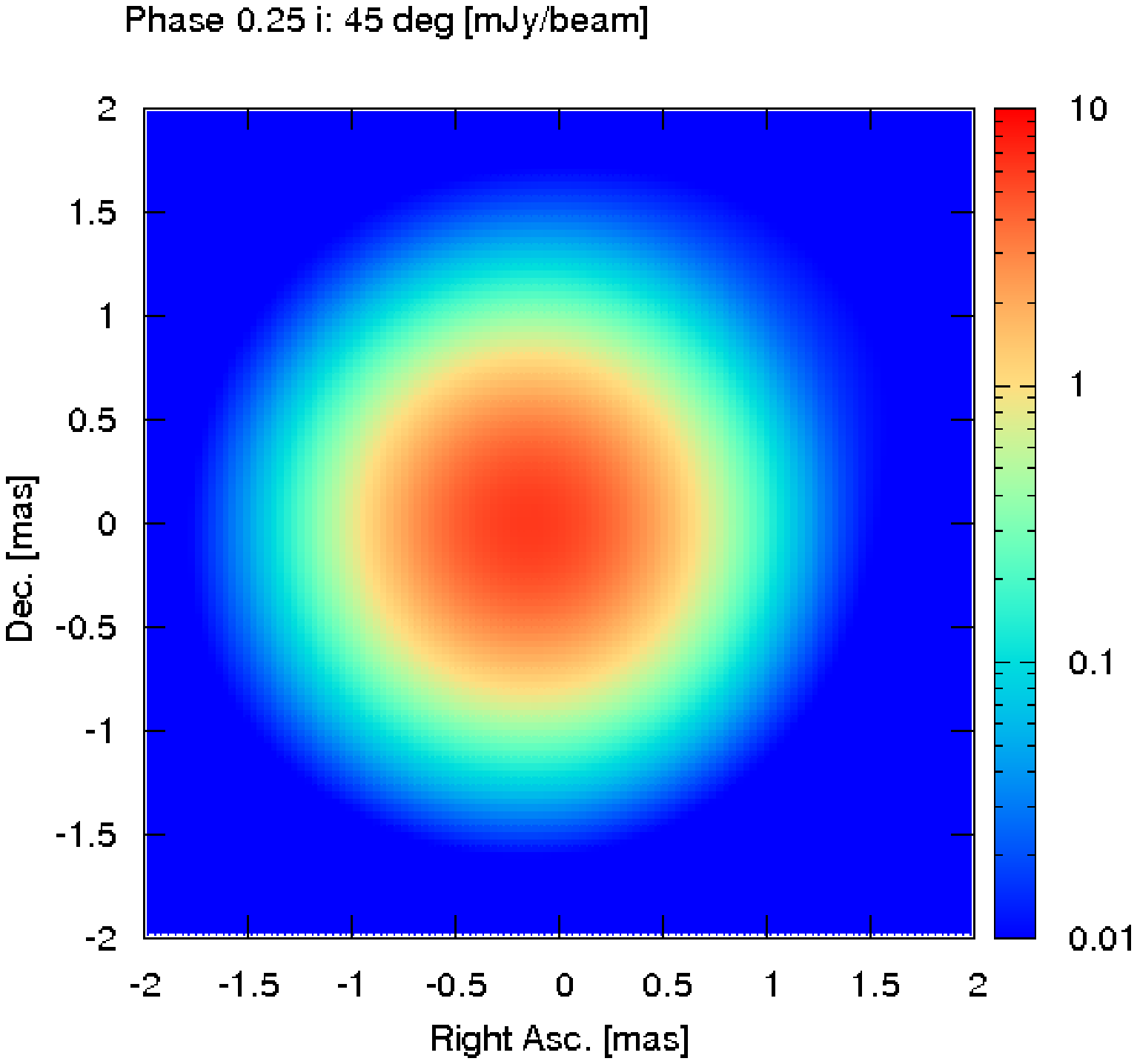}\\[10pt]
\includegraphics[width=0.35\textwidth]{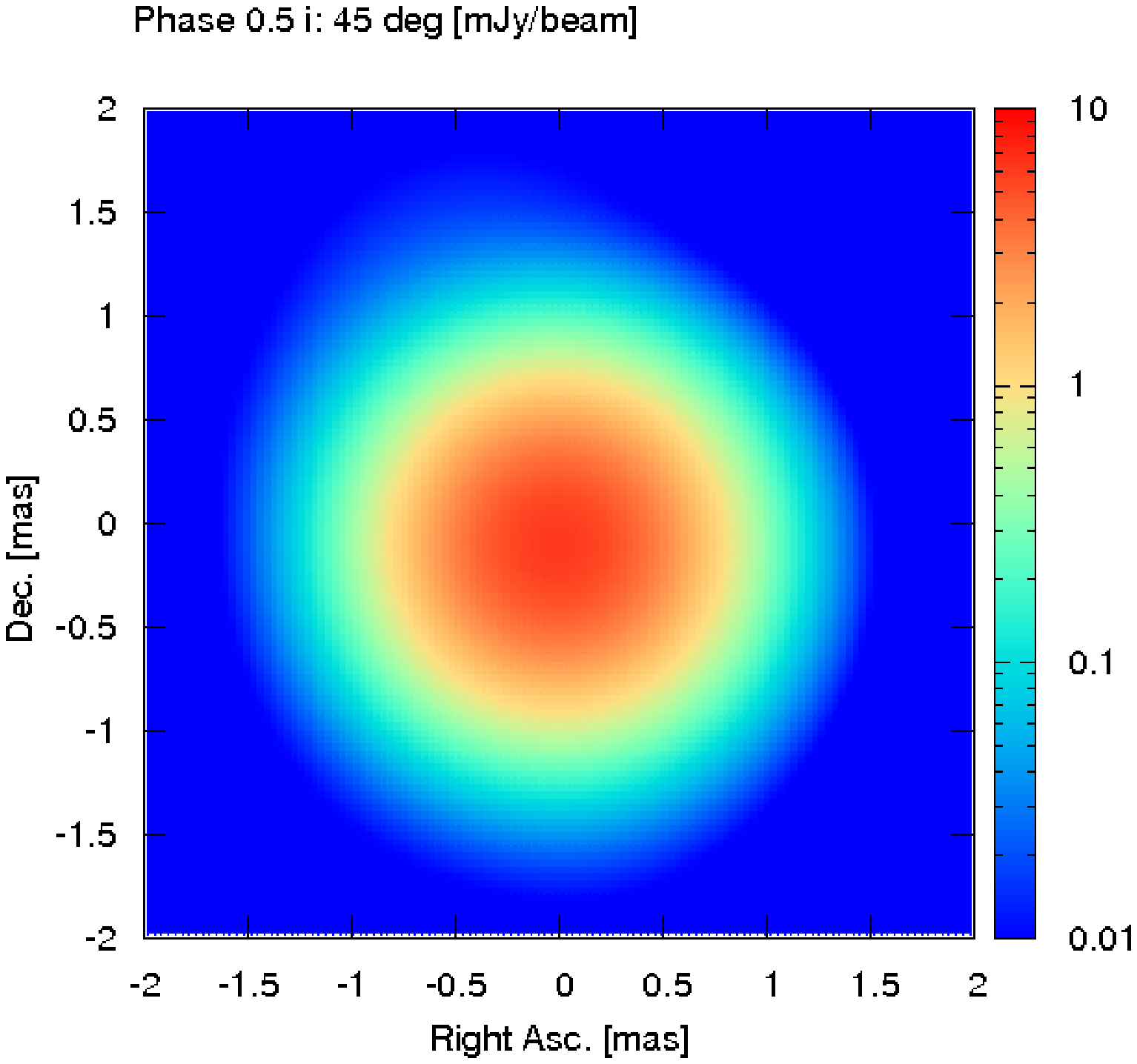}\qquad
\includegraphics[width=0.35\textwidth]{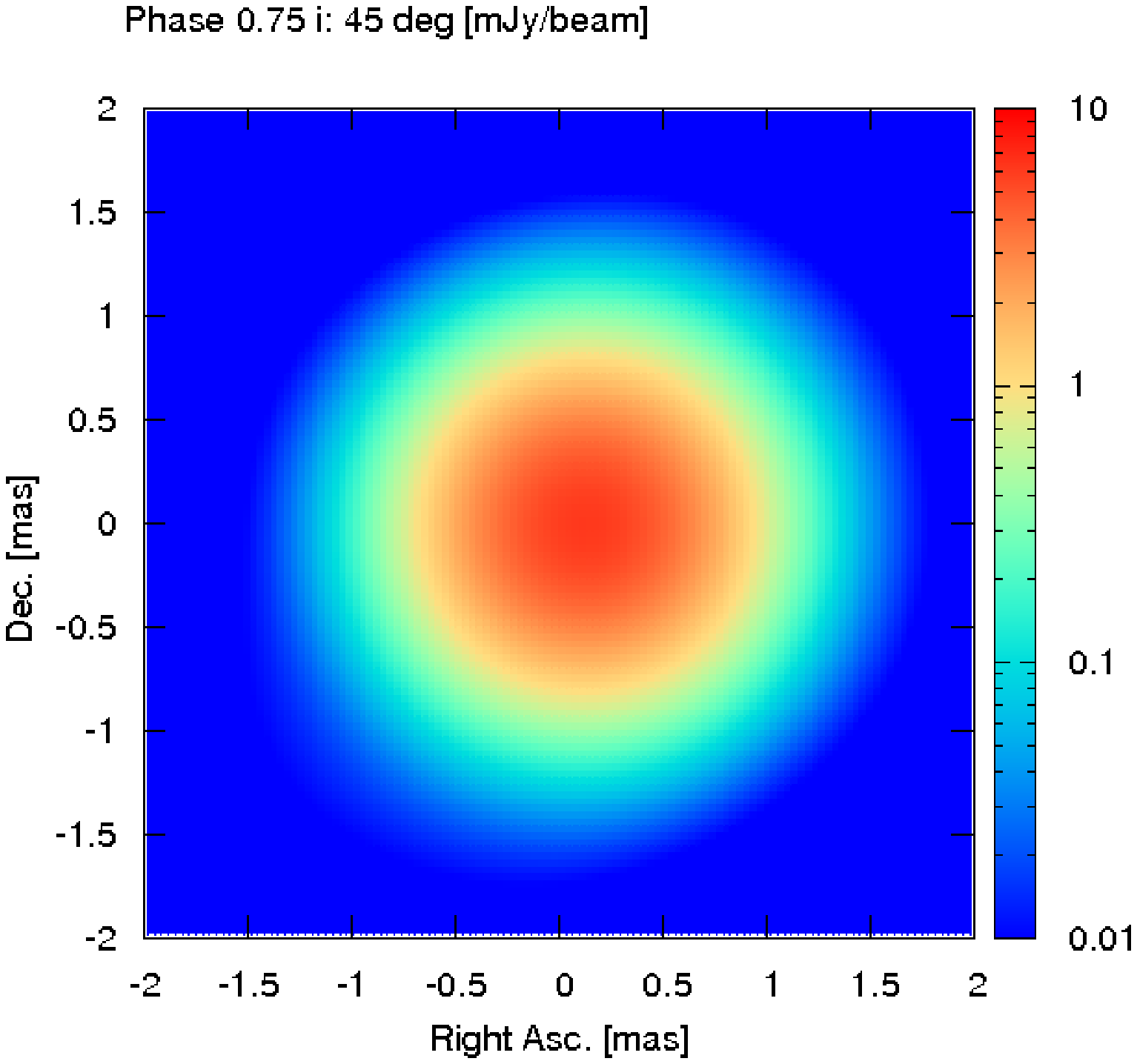}\\
\caption{Gaussian convolved image of the 5~GHz radio emission shown in Fig.~\ref{raw}.}
\label{radd}
\end{figure*}

\begin{figure*}
\centering
\includegraphics[width=0.35\textwidth]{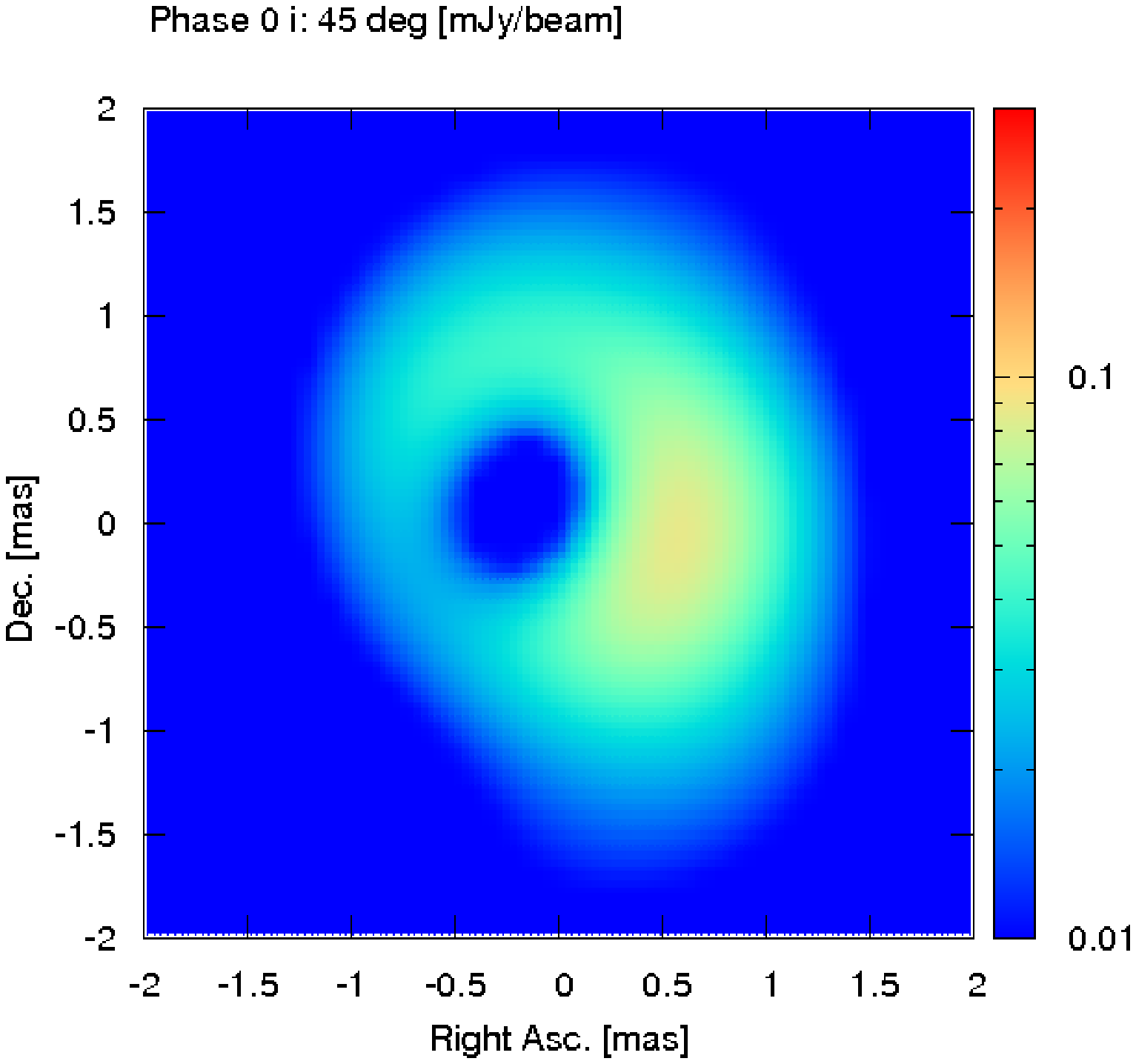}\qquad
\includegraphics[width=0.35\textwidth]{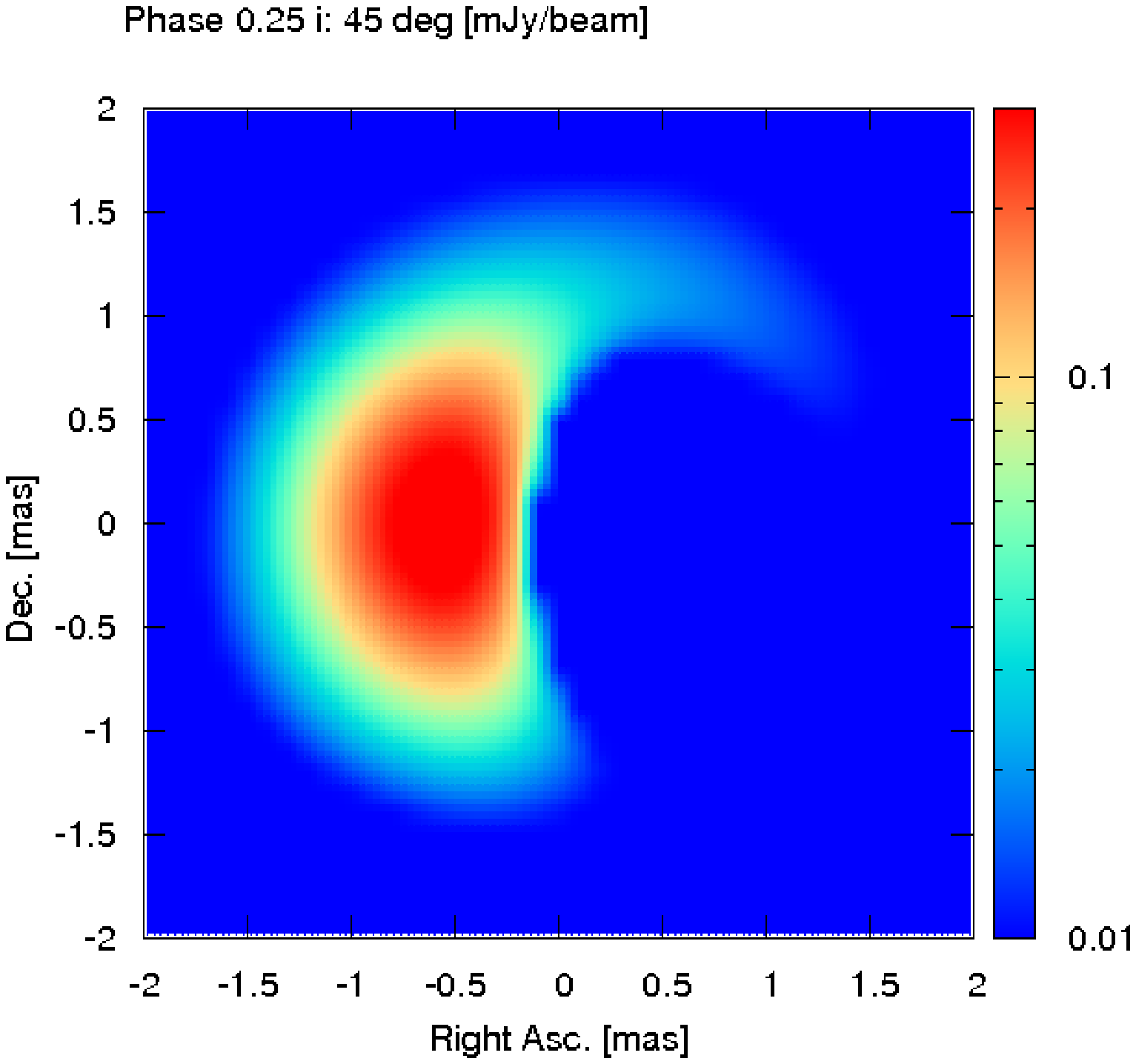}\\[10pt]
\includegraphics[width=0.35\textwidth]{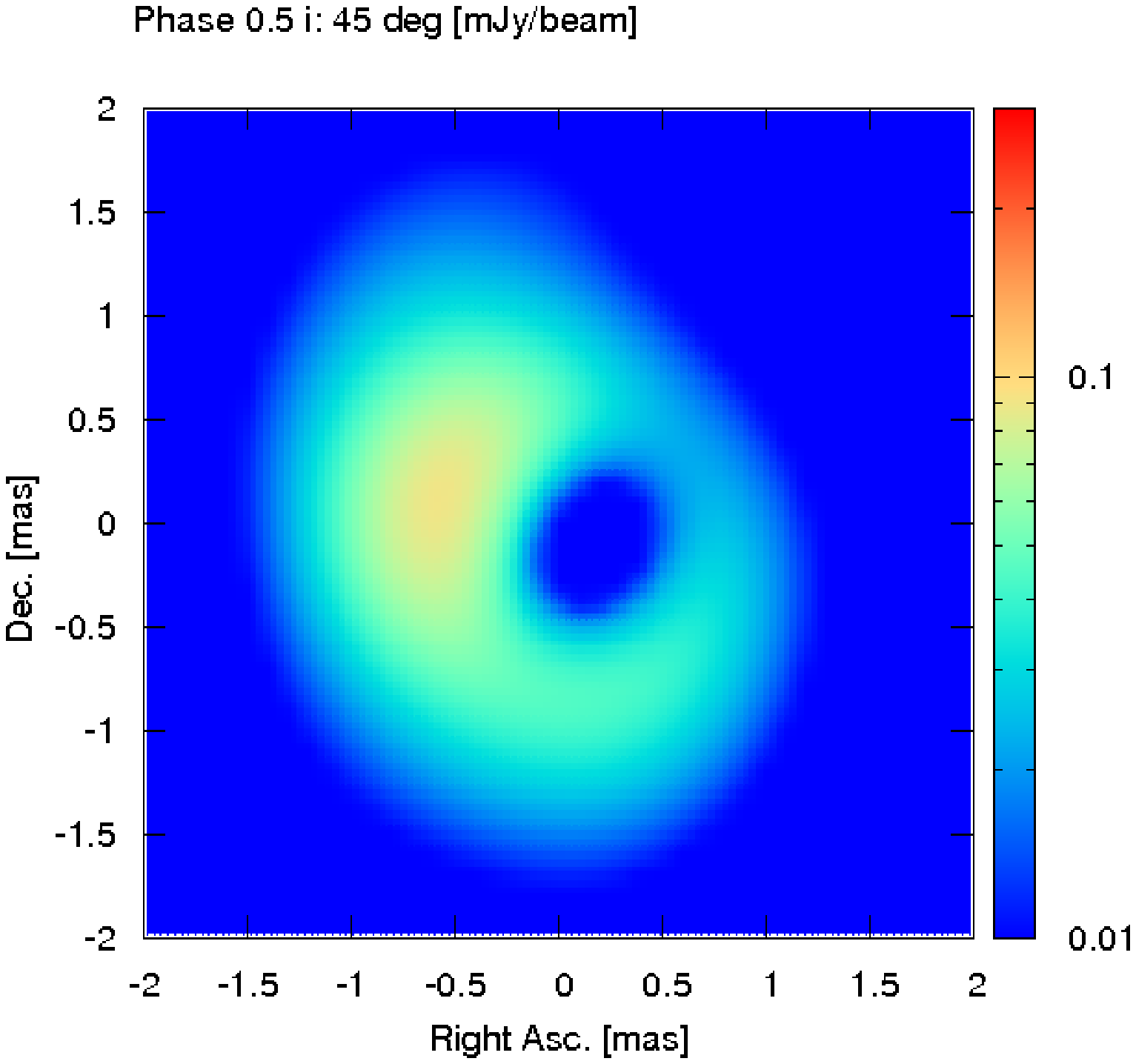}\qquad
\includegraphics[width=0.35\textwidth]{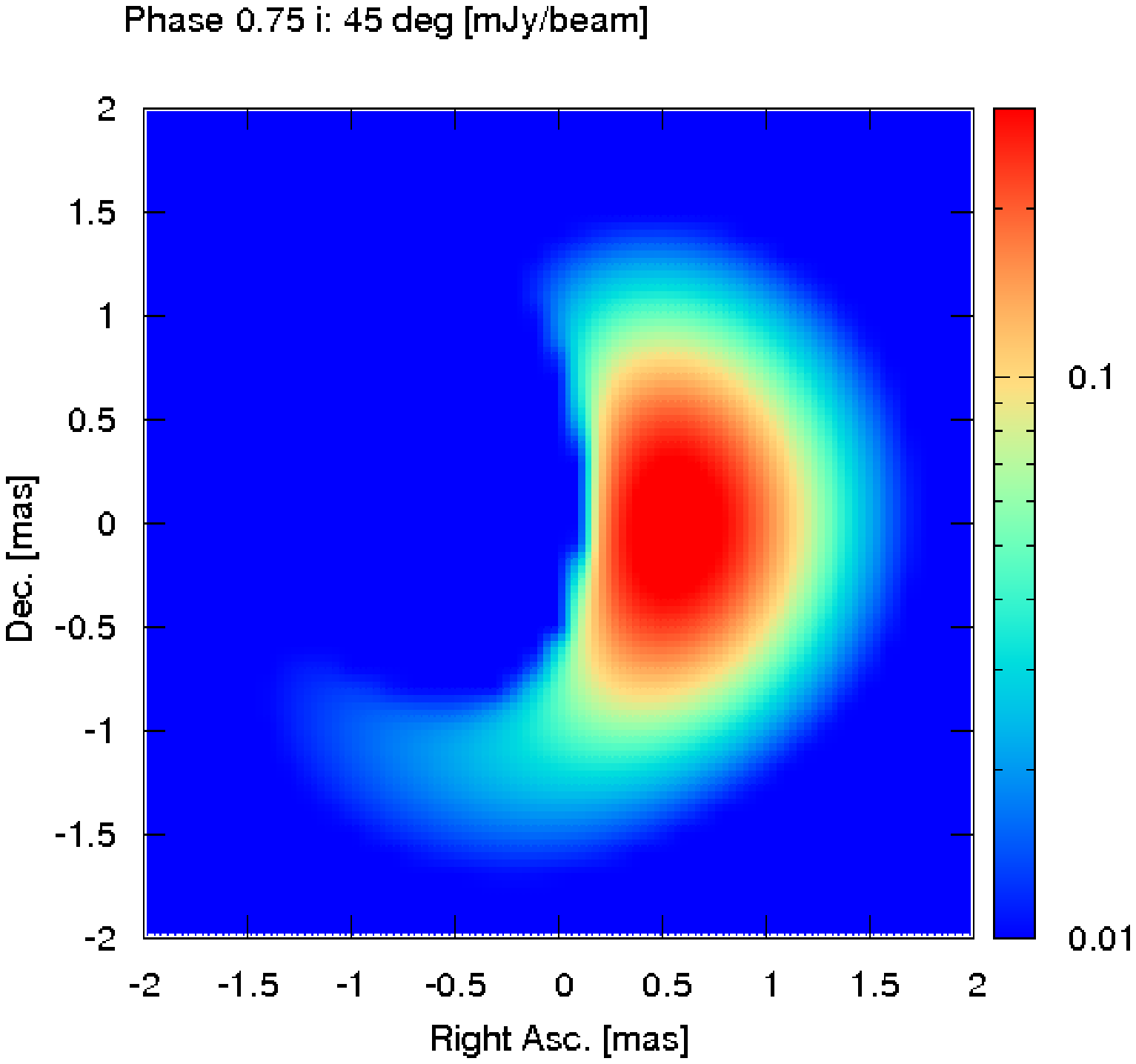}\\
\caption{Residuals of the images presented in Fig.~\ref{radd} after subtracting a point-like (Gaussian) 
source with the same flux.}
\label{er}
\end{figure*}

The radiation image can be computed as it would be seen by the observer projecting the 3D emitter in the  observer
plane and dividing then in 2D bins. From that, synthetic radio maps can be produced convolving the fluxes in these
bins with a circular Gaussian of full width half maximum (FWHM) of 1~mas. The result will be similar to an image obtained by a radio
interferometer with angular resolution of 1~mas.  In Figs.~\ref{raw}, \ref{radd} and \ref{er}, we show three sets of 5~GHz
radio images at four different phases: 0.0 (supc), 0.25, 0.5 (infc), and 0.75. Fig.~\ref{raw} corresponds to the
raw image, i.e. without Gaussian convolution.  There is an expectable similarity in shape between the  radio
secondary (Fig.~\ref{evol2}) and the  flux spatial distribution. The image convolved with a Gaussian with
FWHM=1~mas is presented in Fig.~\ref{radd}, and Fig.~\ref{er} shows the residuals when extracting a point-like source (the
FWHM=1~mas Gaussian)  to the convolved image both with the same flux. Typical noise levels for such an 
interferometer could be $\sim 0.05-0.1$~mJy/beam, slightly below the predicted excesses (see the intensity scale
in Fig.~\ref{er}). The impact of free-free absorption has not been included in these plots, which would consist on
a reduction of the fluxes in the image core but keeping an excess in the periphery.

We have focused in this work on the synchrotron radio emission. We note nevertheless that for the parameter values
adopted here, the synchrotron X-ray and IC GeV fluxes would be both about few $10^{-11}$~erg~s$^{-1}$~cm$^{-2}$. A
more detailed study of the higher energy secondary radiation will be presented elsewhere.

\section{Application to LS~5039}

We have applied our Monte-Carlo code and radiation model presented above to LS~5039. LS~5039 is a high-mass X-ray
binary, detected in VHE gamma rays \citep{aha05b,aha06}, which has been observed in radio with different VLBI
instruments for more than a decade \citep{marti98,par00,par02,ribo08}. The work by \cite{ribo08}
presents VLBA observations at 5~GHz in which the radio core appears as marginally resolved, with two small dipolar
elongated extensions that show a moderate change in the position angle. The core and extension fluxes are 18--20~mJy
and 6--4~mJy, respectively, having the whole structure an angular size of several mas. The radio structures found in
the source have been associated either to a jet \citep{par00} or a shocked pulsar wind \citep{dubus06b}. The binary
system consists of an O6.5 main sequence star and a compact object of unclear nature (neutron star or black hole;
see \cite{cas05}). The system semi-major axis is $a=2.2\times 10^{12}$~cm, with eccentricity $e=0.34$ and orbital
period $P=3.9$~days. The phase 0.0 corresponds to the periastron passage, and phases 0.05 and 0.67 to supc and infc
of the compact object, respectively. The inclination angle of the orbit is not known and could be in a range between
$i\approx 15^\circ-75^\circ$. The stellar mass-loss rate has been taken to be $\sim 3\times 10^{-7}\,M_{\odot}/$yr.
The ephemeris, orbital parameters, system-observer geometry, star and stellar wind properties, and inclination angle
constraints have been taken from \cite{cas05} and \cite{ara09} (see also \cite{sar10}). The system orientation in
the plane of the sky, unconstrained from observations, has been taken as in Fig.~6 of \cite{ara09}. Following the
approach for the generic case, $i$ has been fixed to $45^\circ$ (though our results should not change qualitatively
for $i$ values $45\pm 15^\circ$). The values of $B_*$ and $\eta$ have been taken also as in the generic case, 200~G
and 1, respectively. The parameters adopted for LS~5039 are presented in Table~2.

Given the uncertainties in the source properties, and the known complexities of the orbit and VHE flux/photon-index
lightcurve, we do not aim in this section at carrying a thorough analysis of the different possibilities, what is
left for future work. Otherwise, we intend to give two possible instances of the source behavior, and for that we
focus on two phases, 0.2, slightly after supc, and 0.8, slightly after infc. The distribution of the gamma rays
injected in the system has been chosen such that the source should show absorbed fluxes and photon indices similar
to the observed ones \citep{aha06} in the direction to the observer. For this, we have deabsorbed the observed VHE
spectra (see, e.g., \cite{boettcher07}) assuming a point-like emitter in a certain location, as explained below,
emitting in the energy range from 20~GeV up to 2~TeV. Since the observed lightcurve shows a relatively smooth
behavior around phases 0.45--0.9 and 0.9--0.45 \citep{aha06}, for simplicity we have adopted typical specific fluxes
and photon indices to compute the radio emission:  $n_{\rm TeV}=10^{-12}$~TeV$^{-1}$~s$^{-1}$~cm$^{-2}$ and
$\Gamma=2.5$ (phase 0.2), and $3\times 10^{-12}$~TeV$^{-1}$~s$^{-1}$~cm$^{-2}$ and $\Gamma=2$ (phase 0.8). Since the location of
the emitter is not well constrained \citep{khangulyan08,bosch08b}, two different heights have been chosen that
optimize the radio fluxes without entering in strong conflict with the X-ray fluxes. At phase 0.2, we have located
the emitter at a height of $10^{12}$~cm, and at $2\times 10^{12}$~cm at phase 0.8, { perpendicularly to the orbital 
plane, in the observer side of that plane}. There is no additional emitter assumed at a location symmetric with respect to the
orbital plane, although if gamma-rays were produced in a symmetric bipolar jet, this component should be also
{ included in the calculations. However, given the expected higher wind free-free opacity for this component, 
its contribution is expected to be smaller than the one considered here.}
We have computed the secondary injection and evolution starting a phase interval of 0.5 earlier than the
phases in which the radiation is evaluated, i.e. $0.7\rightarrow 0.2$ and $0.3\rightarrow0.8$. However, most of the
radio emission, including the outermost region of the emitter, comes from secondary pairs injected within a phase
interval of 0.2 earlier, due to adiabatic cooling.  

\begin{figure*}
\centering
\includegraphics[width=0.35\textwidth]{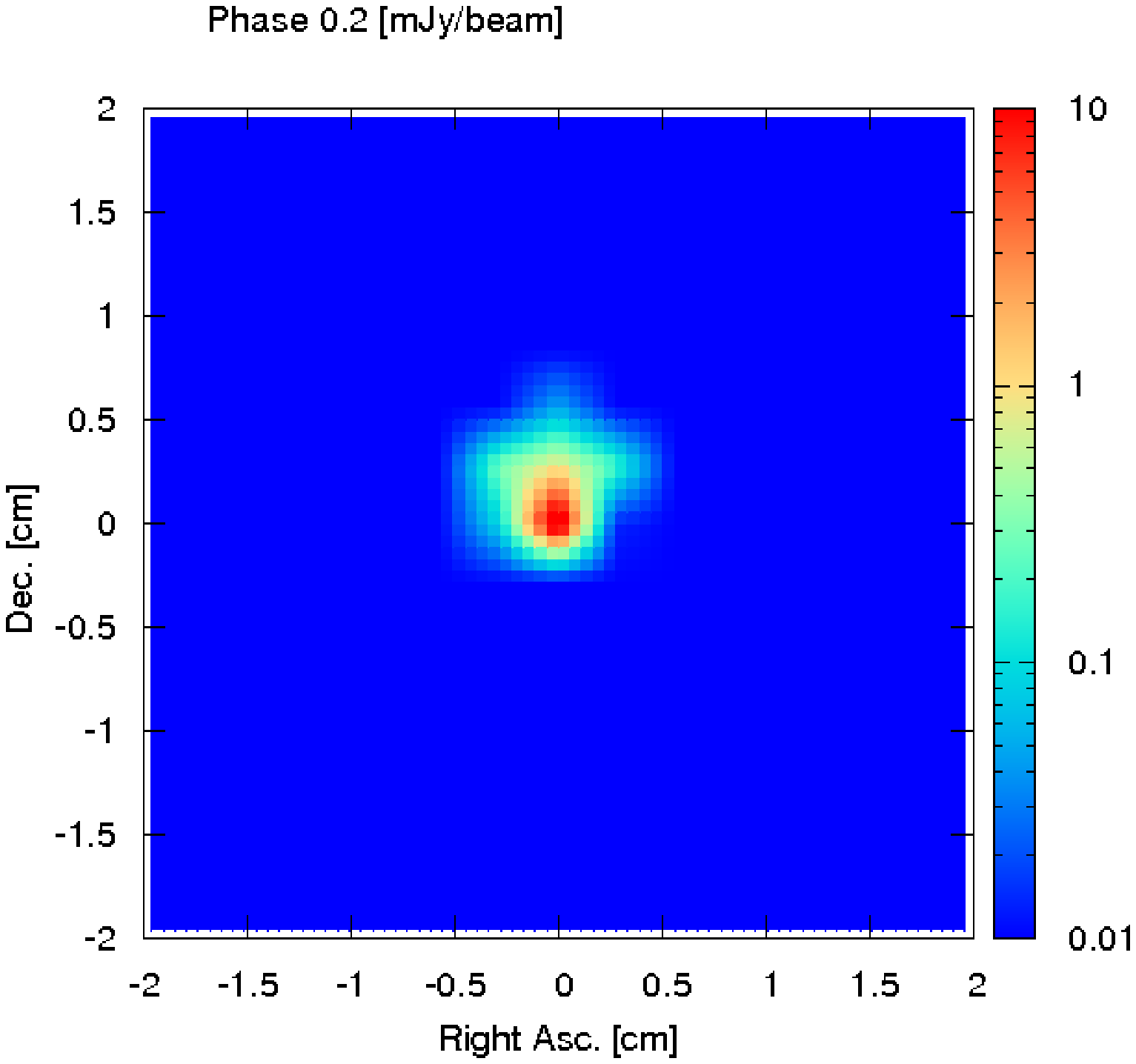}\qquad
\includegraphics[width=0.35\textwidth]{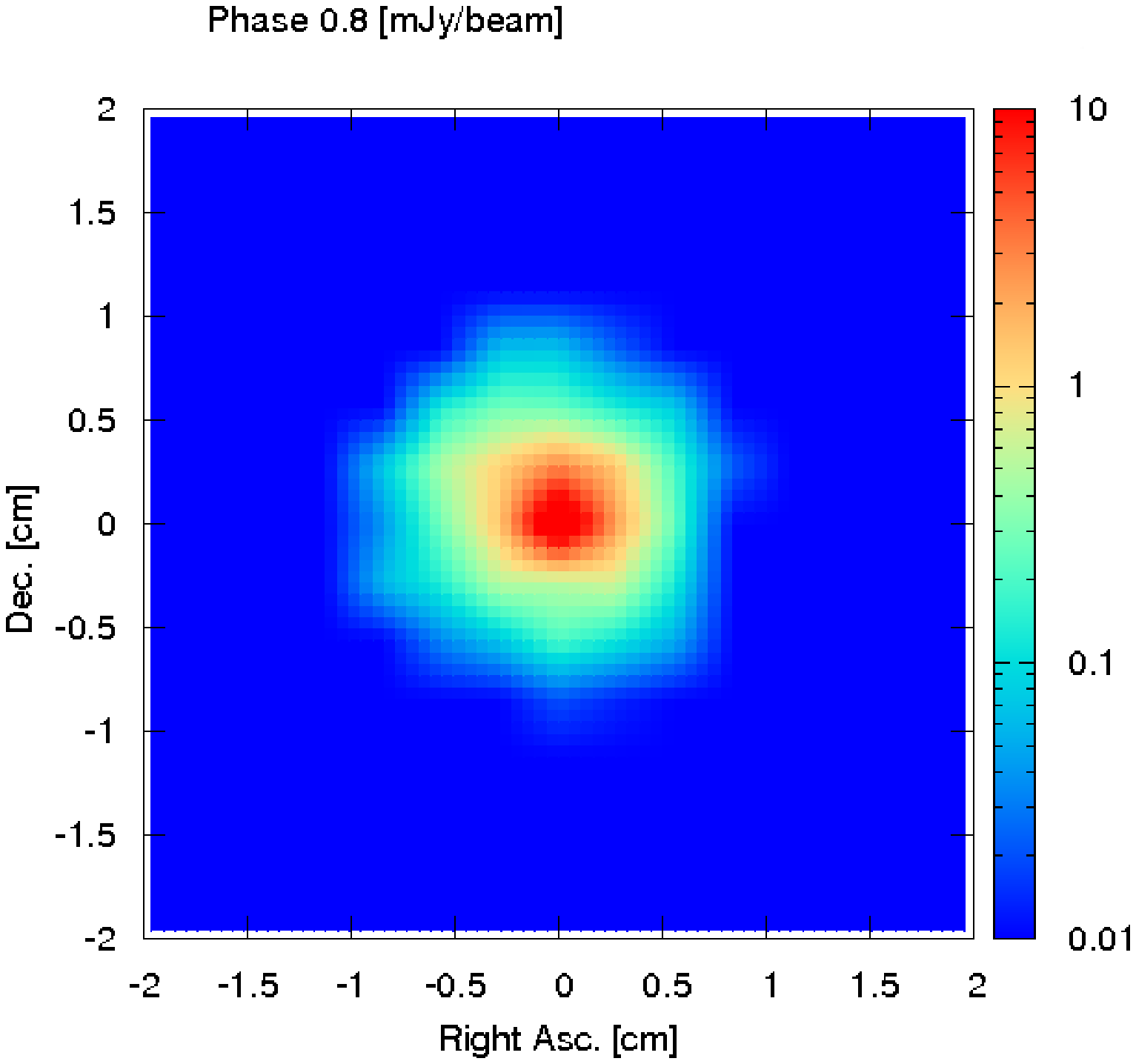}\\[10pt]
\includegraphics[width=0.35\textwidth]{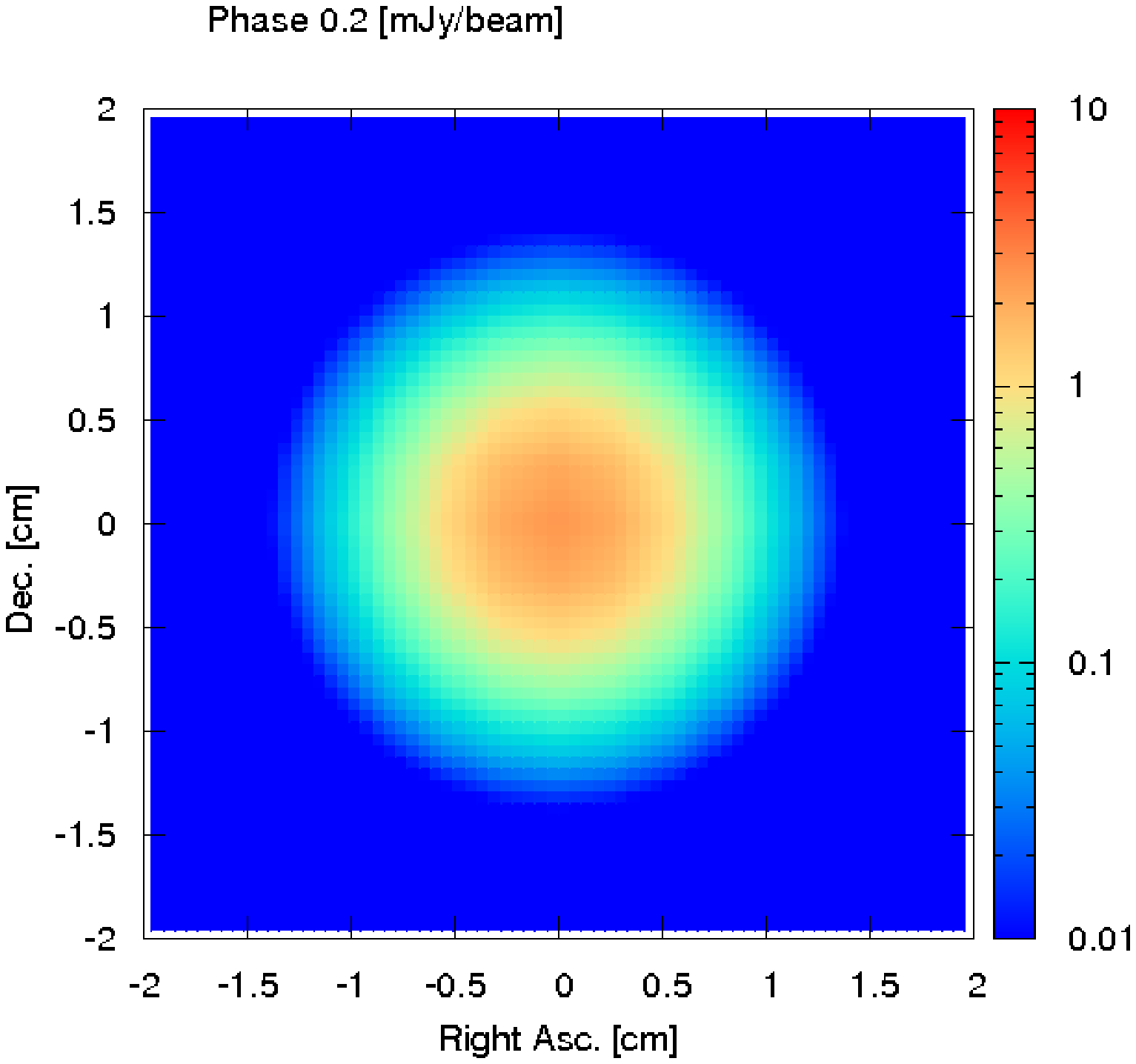}\qquad
\includegraphics[width=0.35\textwidth]{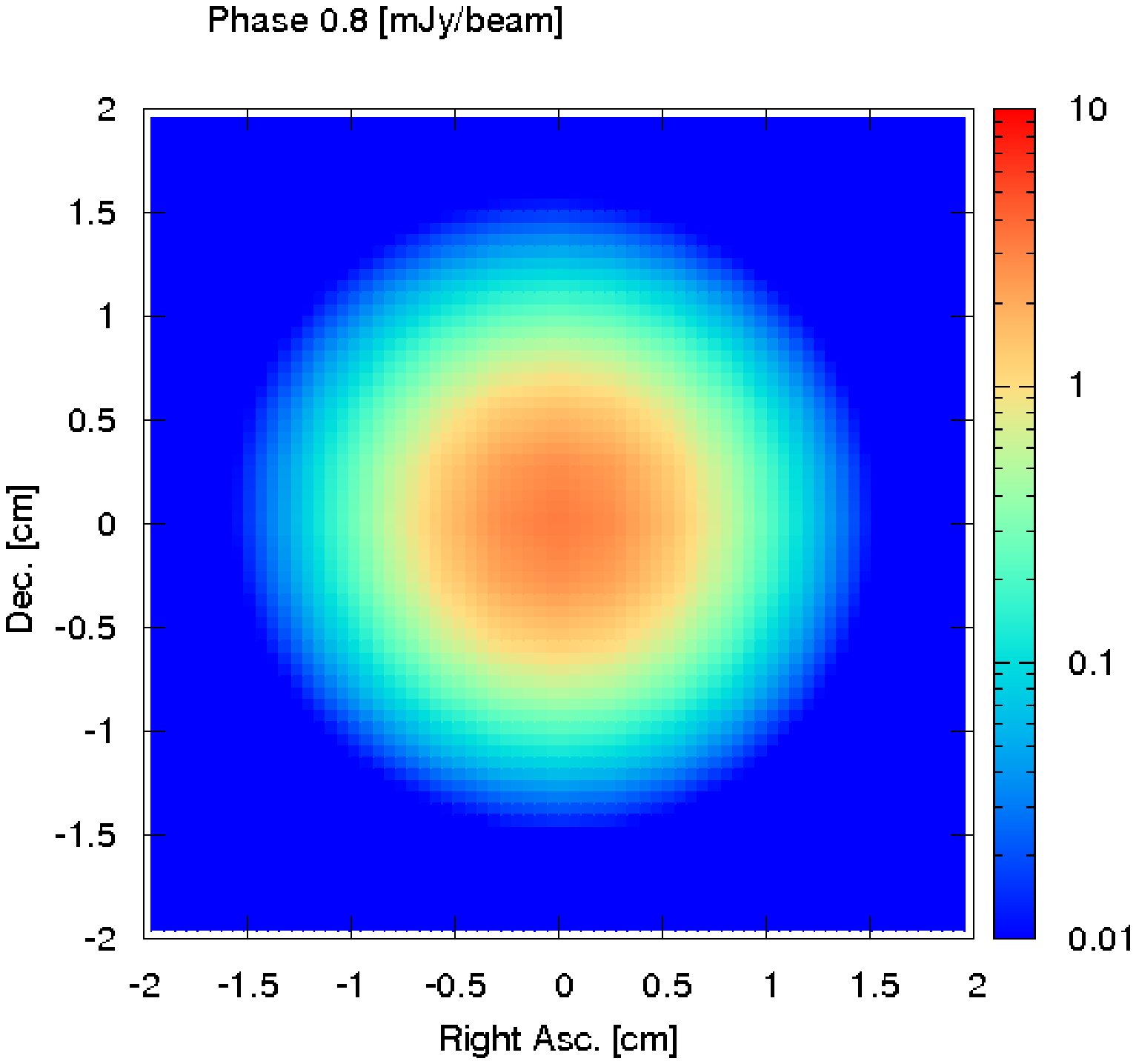}\\
\includegraphics[width=0.35\textwidth]{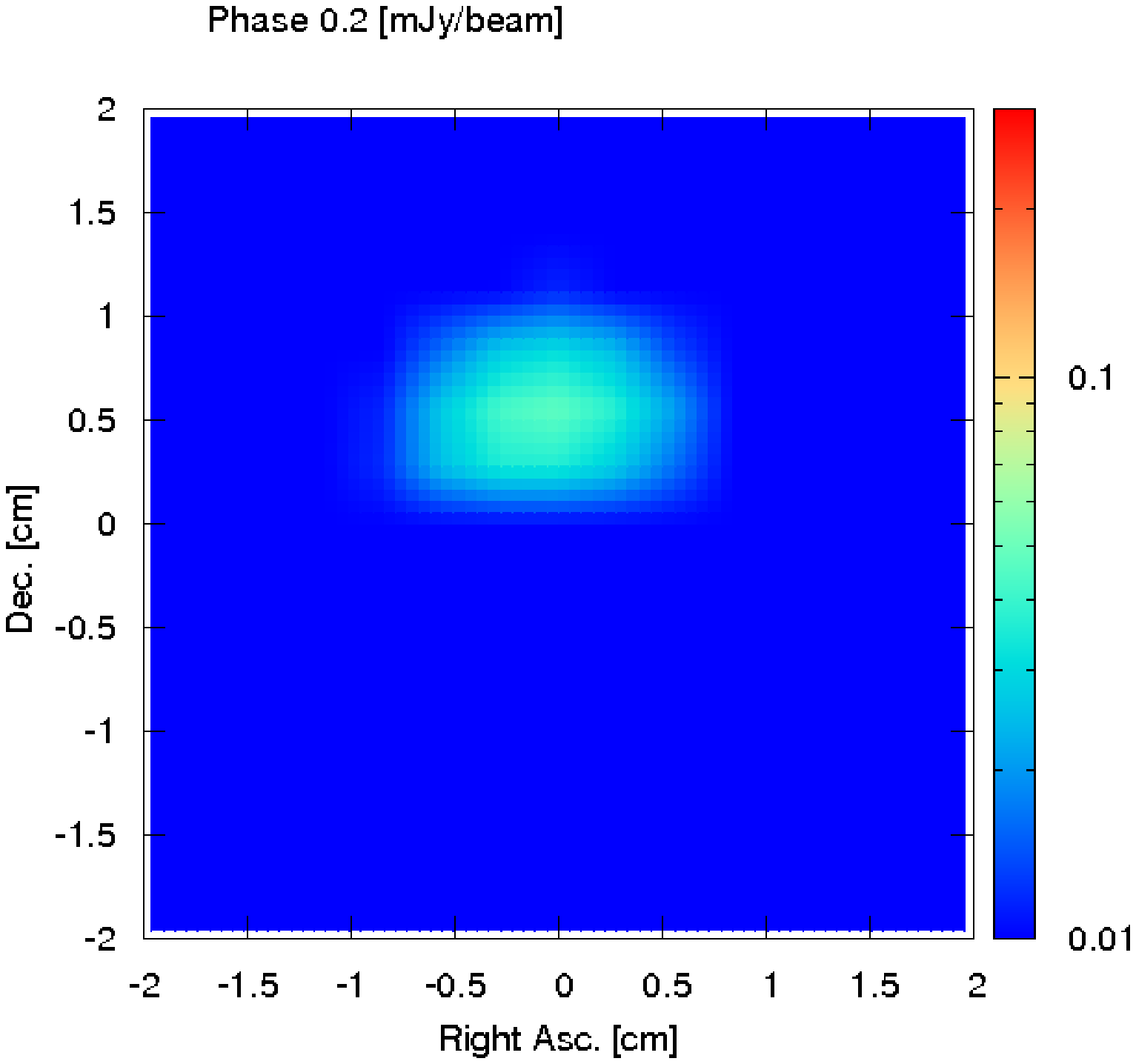}\qquad
\includegraphics[width=0.35\textwidth]{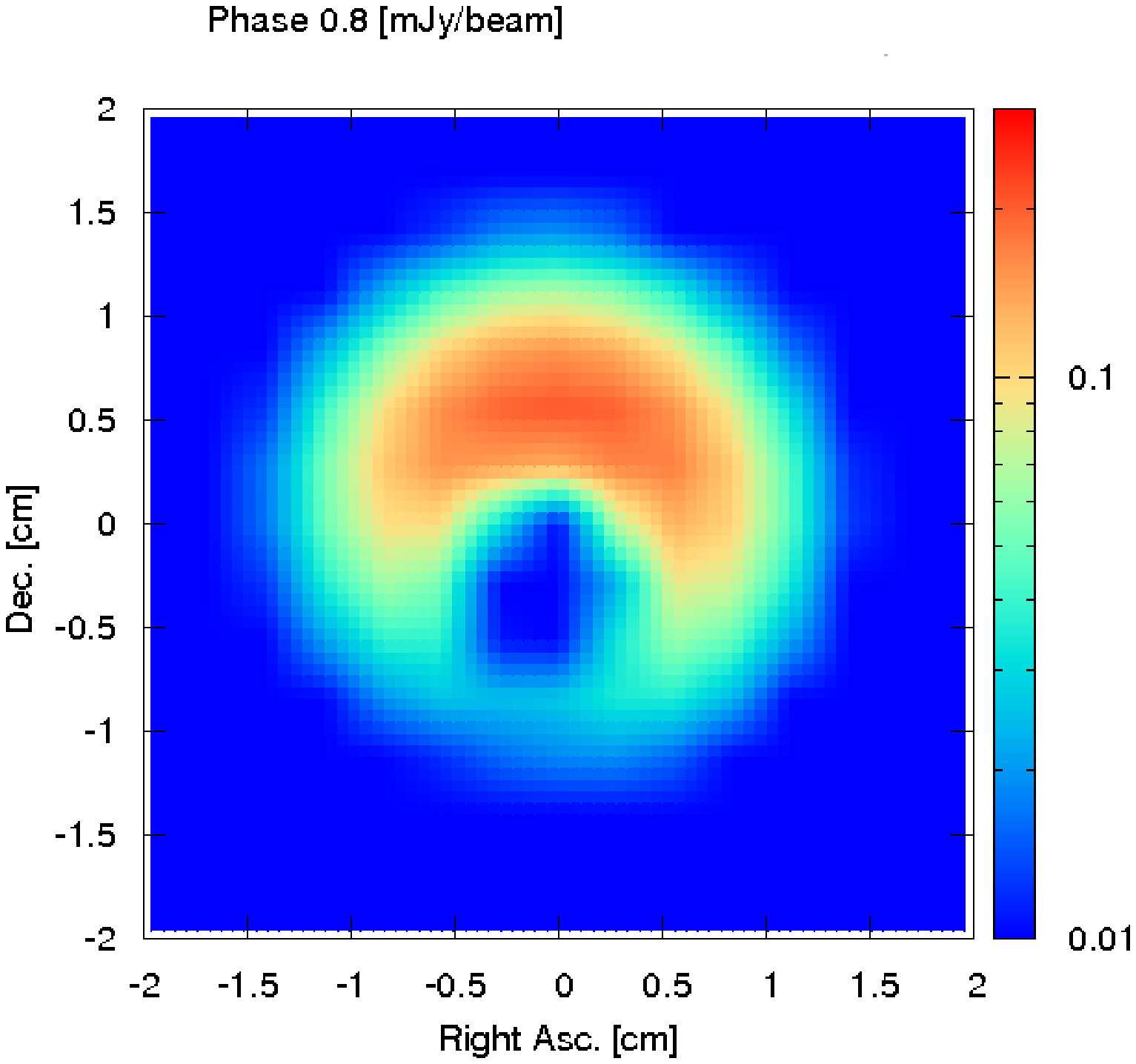}\\
\caption{Computed raw (top), Gaussian convolved (middle), and residual images (bottom) 
of the emission at 5~GHz from LS~5039 in the direction to the observer for the 
phases 0.2 (right) and 0.8 (left).}
\label{ls}
\end{figure*}

The resulting images of the raw radio emission, of the source convolved with a Gaussian, and the residuals after
extracting a point-like source are presented in Fig.~\ref{ls}. The figure shows that the radiation, with a typical
emitter size of $\sim 1$~mas, would be difficult to resolve, although the bigger size of the system around infc may
lead to a marginal extension. The obtained total fluxes at 5~GHz are about 6 and 10~mJy at phases 0.2 and 0.8,
respectively. The complex flux evolution along the orbit of the primary (deabsorbed) gamma-ray spectra yield
different spectra from those of the generic case, obtaining $F_\nu\propto \nu^{-0.5}$ (0.2) and $\propto \nu^{-1}$
(0.8). This is due to secondary pairs injected at different phases, with different absorbed luminosities, produce
the radiation at different frequencies at the same orbital phase. The computed luminosities in the range 1--10~keV
are of about $10^{34}$~erg~s$^{-1}$, similar to those observed in this energy range (e.g. \cite{bosch07,tak09}). These
X-ray fluxes already constrain the fluxes of the radio emission to several mJy at most. These results are
nevertheless sensitive to the radial profile of $B_{\rm w}$, which is not well known, thus we do not treat here the
high-energy output of the secondary emission. The impact of free-free absorption in the wind has not been considered
either, but this process could reduce significantly the secondary radio emission mainly for phases around supc under
high ionization fractions.

 \begin{table}[]
  \begin{center}
  \caption[]{LS~5039 properties}
  \label{tabls}
  \begin{tabular}{lll}
  \hline\noalign{\smallskip}
  \hline\noalign{\smallskip}
Parameter [units] & Symbol  &  Value   \\
  \hline\noalign{\smallskip}
Stellar radius [cm] & $R_*$ & $7\times 10^{11}$ \\
Orbital semi-major axis [cm] & $R_{\rm orb}$ & $2.2\times 10^{12}$ \\
eccentricity & $e$ & 0.34 \\
Stellar temperature [K] & $T_*$ & $3.8\times 10^4$ \\
total system mass [$M_{\odot}$] & $M_*+M_{\rm X}$ & 25 \\
Star surface magnetic field [G] & $B_*$ & 200 \\
Mass-loss rate [$M_{\odot}$~yr$^{-1}$] & $\dot{M}$ & $3\times 10^{-7}$ \\
Wind speed at infinity [cm~s$^{-1}$] & $v_{\infty}$ & $2.4\times 10^8$ \\
Distance [kpc] & $d$ & 2.5 \\
Superior conjunction & supc & 0.05 \\
Inferior conjunction & infc & 0.67 \\
Inclination angle [$^\circ$] & $i$ & 45 \\
Irregular magnetic field fraction & $\eta$ & 1 \\
  \noalign{\smallskip}\hline
  \end{tabular}
  \end{center}
\end{table}

\section{Discussion}

This work shows that the radio emission detected from compact gamma-ray binaries { that contain} a massive star could
have a non-negligible component coming from the secondary pairs created in the vicinity of the gamma-ray emitter. 
Actually, the following simple estimate of the radio flux already indicates this. Assuming that secondary pairs
cool mainly through adiabatic cooling in the region in which they produce radio emission, which is the case for a
wide range of parameter values, and a wind velocity of $2\times 10^8$~cm~s$^{-1}$, the expected fluxes can be
derived from the formula:
\begin{eqnarray}
F_{\rm 5~GHz}&\sim&(1/4\pi d^2)\,L_{e^\pm}\,(E_{\rm radio}/E_{\rm 0})\,(t_{\rm ad}/t_{\rm sync})\,\tau_{\rm 100}
\\ {} & = &
20
\,(L_{\rm VHE}/10^{35}\,{\rm erg~s}^{-1})
\,(B_{\rm e\phi}/10\,{\rm G})
\nonumber
\label{rad}
\end{eqnarray}
$$
\times (L_*/10^{38}\,{\rm erg~s}^{-1})
\,(R_{\rm e}/3\times 10^{12}\,{\rm cm})^{-1}
\,(d/2\,{\rm kpc})^{-2}
\,{\rm mJy}\,,
$$
where $E_{\rm radio}$ is the energy of the radio emitting particles, $\tau_{\rm 100}$ the angle averaged
opacity at gamma-ray energies around 100~GeV, $t_{\rm sync}$ the synchrotron cooling timescale,  $E_0\sim 30$~GeV
the secondary typical injection energy, $L_{\rm VHE}$ the VHE luminosity, $B_{\rm e\phi}=B_\phi(R_{\rm e})$, and 
$d$ the distance to the source. In Eq.~(\ref{rad}), the product  $(L_*/10^{38}\,{\rm erg~s}^{-1})\,(R_{\rm
e}/3\times  10^{12}\,{\rm cm})^{-1}$ should be substituted by $\sim 2$ for $\tau_{\rm 100}\ge 1$ (and
$\tau_{\rm 100}$ fixed to 1).  In sources with  parameters such that $F_{\rm 5~GHz}\gtrsim 1$~mJy, secondary radio
emission should not be neglected. The most crucial point here is whether magnetic fields of the order of 10~G can
be found at few $R_*$ from the star.

As shown in this work, secondary radio emission from gamma-ray binaries would not be only relevant in flux, but
they could also present resolvable extensions. In general, the spectra can be flat or softer depending on $u_*$
(adiabatic vs IC cooling dominance), and the absorbed gamma-ray luminosity and the radio emitter conditions, in
particular $v_{\rm w}$ and $B_{\rm w}$, along the orbit. In the particular case of LS~5039, we find
that a dominant fraction of the marginally extended core of LS~5039 \citep{ribo08} could be of secondary origin,
with a changing relatively soft spectrum. Since the computed additional extension appears marginal, the observed elongated
emission at 5~GHz beyond the core in LS~5039 would be likely related to an intrinsic primary radio emitter.

Regarding morphology, we find that a spiral-like shape would be expected, although typically the small angular size
of the radio source presently prevents detailed imaging. Such a spiral-like shape was also predicted in
\citet{bosch08a}, although the treatment there was simpler than in the present work. It is worthy noting that the
wind conditions could differ from those adopted here, and then the shape, and even the radio emitter extension, may
change. More collimated radio structures, or a magnetic field increase downstream the wind leading to more extended
radio emission, cannot be discarded.

It is remarkable that the spatial distribution of the radio emitting secondary pairs is strongly determined by their
injection, which depends on the stellar photon field and the gamma-ray emitter location, and by the stellar wind
inhomogeneities and anisotropies, but not by $B_{\rm w}$. Only a very regular $B_{\rm w}$, with very large
$\eta$-values, could modify significantly the distribution of radio emitting secondary pairs, since then these
particles would move at $c$ along the dominantly toroidal $B_{\rm w}$, i.e. leaving the system with a speed 
$c\,B_{\rm wr}/B_{\rm w\phi}$, between $v_{\rm w}$ and $c$. For reasonable $B_{\rm w}$ and $\eta$-values, particles
are trapped in the wind, and their energy distribution is determined by IC and adiabatic cooling. The magnetic field
strength is then relevant only for the synchrotron fluxes (recall $F\propto B_{\rm w}$), and its level of
(ir)regularity, $\eta$, for variability. Polarization would also be affected by $\eta$, and polarization
observations may probe the $B_{\rm w}$ geometry. However, the high density of cold free electrons in the stellar
wind could induce strong Faraday rotation. A proper study of the impact of this on the final polarization degree and
angle requires a devoted investigation.

System eccentricity, a significant regular magnetic field component, and free-free absorption would be sources of
radio variability and should be considered when studying specific cases. Changes in the gamma-ray luminosity would
also lead to a smooth modulation of the secondary radio emission. To finish with, it seems likely that in gamma-ray
binaries, the radio emission from inner regions can be, if not dominated, significantly contaminated by secondary
radiation. Secondary pairs should not be neglected when understanding the broadband non-thermal emission in compact
gamma-ray sources, nor the physical connection of the radiation produced in different bands.

\bigskip

We thank an anonymous referee for her/his many constructive and useful comments and
suggestions. The research leading to these results has received funding from the
European Union Seventh Framework Programme (FP7/2007-2013) under grant agreement
PIEF-GA-2009-252463. V.B-R. acknowledges support by the Ministerio de
Educaci\'on y Ciencia (Spain) under grant AYA 2007-68034-C03-01,
AYA2010-21782-C03-01 and FPA2010-22056-C06-02. V.B-R. thanks Max Planck Institut
fuer Kernphysik for its kind hospitality  and support. V.B-R. wants to thank the
Insituto Argentino de Astronom\'ia and the Facultad de Ciencias Astron\'omicas y
Geof\'isicas de la Universidad de La Plata for their kind hospitality.

{}

\begin{thebibliography}{}

\bibitem[Aharonian et al.(2005a)]{aha05a} 
Aharonian, et al. 2005a, A\&A, 442, 1
\bibitem[Aharonian et al.(2005b)]{aha05b} 
Aharonian, et al. 2005b, Science, 309, 746 
\bibitem[Aharonian et al.(2006)]{aha06} 
Aharonian, F.~A. et al. 2006, A\&A, 460, 743
\bibitem[Aharonian et al.(2007)]{aha07} 
Aharonian, F.~A. et al. 2007, A\&A, 469, L1
\bibitem[Albert et al.(2006)]{albert06} 
Albert, J. et al. 2006, Science, 312, 1771
\bibitem[Albert et al.(2007)]{albert07} 
Albert, J. et al. 2007, ApJ, 665, L51
\bibitem[Albert et al.(2009)]{albert09} 
Albert, J. et al. 2009, ApJ, 693, 303
\bibitem[Aragona et al.(2009)]{ara09}
Aragona, C., McSwain, M.~V., Grundstrom, E.~D., et al. 2009, ApJ, 698, 514
\bibitem[Bednarek(2000)]{bed00} 
Bednarek, W. 2000, A\&A, 362, 646 
\bibitem[Boetcher(2007)]{boettcher07}
B\"ottcher, M. 2007, Astr. Phys., 27, 278
\bibitem[Boetcher \& Dermer(2005)]{boettcher05}
B\"ottcher, M., Dermer, C.~D. 2005, ApJ, 634, L81
\bibitem[Bogovalov et al.(2008)]{bogovalov08}
Bogovalov, S.~V., Khangulyan, D.~V., Koldoba, A.~V., Ustyugova,
G.~V., \& Aharonian, F. A. 2008, MNRAS, 387, 63
\bibitem[Bosch-Ramon et al.(2007)]{bosch07} 
Bosch-Ramon, V., Motch, C., Rib\'o, M. 2007, A\&A, 473, 545
\bibitem[Bosch-Ramon et al.(2008a)]{bosch08a} 
Bosch-Ramon, V., Khangulyan, D., Aharonian, F.~A. 2008a, A\&A, 482, 397
\bibitem[Bosch-Ramon et al.(2008b)]{bosch08b} 
Bosch-Ramon, V., Khangulyan, D., Aharonian, F.~A. 2008b, A\&A, 489, L21
\bibitem[Bosch-Ramon(2009)]{bosch09a} 
Bosch-Ramon, V. 2009, A\&A, 493, 829
\bibitem[Bosch-Ramon \& Khangulyan(2009)]{bosch09b} 
Bosch-Ramon, V. \& Khangulyan, D. 2009, Int. Jour. Mod. Phys. D, 18, 347 [astro-ph/0805.4123]
\bibitem[Casares et al.(2005)]{cas05}
Casares, J., Rib\'o, M., Ribas, I., et al. 2005, MNRAS, 364, 899
\bibitem[Corbet et al.(2011)]{cor11} Corbet, R.~H.~D., Cheung, C.~C., Kerr, M., 2011, ATel, 3221, 1
\bibitem[de Ona Wilhelmi(2010)]{deo10} de Ona Wilhelmi, E. for the HESS collaboration, 2010, 38th COSPAR Scientific Assembly, Bremen, Germany, p.6
\bibitem[Dhawan et al.(2006)]{dhawan06}
Dhawan, V., Mioduszewski, A., \& Rupen, M. 2006, in Proc. of the VI Microquasar Workshop, Como-2006 
\bibitem[Dubus(2006a)]{dubus06a}
Dubus, G. 2006, A\&A, 451, 9
\bibitem[Dubus(2006b)]{dubus06b}
Dubus, G. 2006b, A\&A, 456, 801
\bibitem[Blandford \& Eichler(1987)]{bla87} Blandford, R., Eichler, D., 1987, Phys. Rep., 157, 1
\bibitem[Falcone et al.(2010)]{falcone10}
Falcone, A. D., Grube, J., Hinton, J., et al. 2010, ApJ, 708, L52 
\bibitem[Falcone et al.(2011)]{falcone11}
Falcone, A., Bongiorno, S., Stroh M., Holder, J. 2011, ATel 3152
\bibitem[Ford(1984)]{ford84}
Ford, L.~H. 1984, MNRAS, 211, 559
\bibitem[Gould \& Schr\'eder(1967)]{gould67}
Gould, R.~J. \& Schr\'eder, G.~P. 1967, Phys. Rev., 155, 1404
\bibitem[Hinton et al.(2009)]{hinton09} 
Hinton, J.~A., Skilton, J.~L., Funk, S., et al. 2009, ApJ, 690, L101
\bibitem[Jones(1990)]{jones90} Jones, F. C. 1990, ApJ, 361, 162
\bibitem[Khangulyan et al.(2008)]{khangulyan08}
Khangulyan, D., Aharonian, F., \& Bosch-Ramon, V. 2008, MNRAS, 383, 467 
\bibitem[Mart\'i et al.(1998)]{marti98}
Mart\'i, J., Paredes, J.~M., \& Ribo, M. 1998, A\&A, 338, L71
\bibitem[Mold\'on et al.(2011)]{mol11} Mold\'on, J., Rib\'o, M., Paredes, J.~M. 2011, ATel, 3180, 1
\bibitem[Moskalenko \& Karacula(1994)]{moskalenko94}
Moskalenko I. V., Karakula S., 1994, ApJ, 92, 567
\bibitem[Orellana et al.(2007)]{orellana07}
Orellana, M., Bordas, P., Bosch-Ramon, V., Romero, G. E., \& Paredes, J. M. 2007, A\&A, 476, 9
\bibitem[Paredes et al.(2000)]{par00}
Paredes, J. M., Mart\'i, J., Rib\'o, M., \& Massi, M. 2000, Science, 288, 2340
\bibitem[Paredes et al.(2002)]{par02}
Paredes, J. M., Rib\'o, M., Ros, E., Mart\'i, J., \& Massi, M. 2002, A\&A, 393, L99
\bibitem[Protheroe \& Stanev(1987)]{protheroe87}
Protheroe, R.~J. \& Stanev, T. 1987, ApJ, 322, 838
\bibitem[Reynoso et al.(2008)]{rey08} 
Reynoso, M.~M., Christiansen, H.~R., \& Romero, G. E., 2008, Astrop. Phys. 28, 565
\bibitem[Rib\'o et al.(2008)]{ribo08} 
Rib\'o, M., Paredes, J.~M., Mold\'on, J., Mart\'i, J., Massi, M. 2008, A\&A, 481, 17
\bibitem[Romero et al.(2010)]{rom10} 
Romero, G.~E., del Valle, M.~V., \& Orellana, M. 2010, A\&A, 518, 12
\bibitem[Rybicki \& Lightman(1979)]{ryb79}
Rybicki, G. B., \& Lightman, A. P. 1979, Radiative processes in astrophysics
(New York: Wiley-Interscience)
\bibitem[Sarty et al.(2010)]{sar10}
Sarty, G.~E., Szalai, T., Kiss, L.~L. 2010, MNRAS, in press [astro-ph/1009.5150]
\bibitem[Sierpowska-Bartosik \& Torres(2008)]{sier08} 
Sierpowska-Bartosik, A. \& Torres, D.~F., 2008, APh, 30, 239
\bibitem[Skilton et al.(2009)]{skilton09}     
Skilton, J. L., Pandey-Pommier, M., Hinton, J. A., et al. 2009, MNRAS, 399, 317
\bibitem[Takahashi et al.(2009)]{tak09}
Takahashi, T., Kishishita, T., Uchiyama, Y., et al. 2009, ApJ, 697, 592
\bibitem[Usov \& Melrose(1992)]{usov92} 
Usov, V.~V.,\& Melrose, D.~B. 1992, ApJ, 395, 575

\end{thebibliography}
\end{document}